\documentclass[english,12pt]{article}
\usepackage[cp1251]{inputenc}
\usepackage{babel}
\usepackage{amsfonts, amsmath}

\newcommand{\be}{\begin{equation}} \newcommand{\ee}{\end{equation}}

\newcommand{\gsim}{\mathrel{\hbox{\rlap{\lower.55ex\hbox{$\sim$}} \kern-.3em \raise.4ex \hbox{$>$}}}}

\begin{document}
\begin{center}
{\bf Two Approaches to Measurability Concept  and Quantum
Theory
}\\
\vspace{5mm} Alexander Shalyt-Margolin \footnote{E-mail:
a.shalyt@mail.ru; alexm@hep.by}\\ \vspace{5mm} \textit{Research
Institute for Nuclear Problems, Belarusian State University, 11
Bobruiskaya str., Minsk 220040, Belarus}
\end{center}
 \rm\normalsize \vspace{0.5cm}

\begin{abstract}
In the present paper, in terms of the  measurability concept
introduced in the previous works of the author, a quantum theory
is studied.   Within the framework of this concept, several
examples are considered using the Schrodinger picture; analogs of
Fourier transformations from the momentum representation to the
coordinate one and vice versa are constructed. It is shown how to
produce a measurable analog of the Heisenberg picture. At the end
of this paper the obtained results are used to substantiate
another (more general) definition of the measurability concept
that is not based on the Heisenberg Uncertainty Principle and its
generalization, as it has been in the previous works of the
author, and may be equally suitable for both non-relativistic and
relativistic cases.
\end{abstract}

\section{Introduction.}

The present paper is a continuation of the previous works published by the author on the subject
\cite{Shalyt-AHEP2}--\cite{Shalyt-new1}. The main idea and target
of these works is to construct a correct quantum theory and
gravity in terms of the variations (increments) dependent on the
existent energies.
\\It is clear that such a theory should not involve infinitesimal space-time variations
\begin{equation}\label{Introd.1}
dt,dx_{i},i=1,...,3.
\end{equation}
\\ The main instrument in the above-mentioned articles was the {\bf measurability} concept introduced in\cite{Shalyt-Entropy2016.1}.
\\Within the framework of this concept, a theory becomes discrete
but at low energies $E$ far from the Planck energies
$E\ll E_p$ it becomes very close to the initial theory in
the continuous space-time paradigm.
\\ Quantum mechanics is studied in the present work in terms of the {\bf measurability} notion.
In Sections 2,3 the results earlier obtained by the
author are elaborated as applied to non-relativistic and relativistic
quantum theories.
\\ Section 4 presents a {\bf measurable} analog
of the wave function and the Schrodinger Equation as well as the
main differential operators involved in quantum mechanics, in
particular, the Laplace operator.
\\{\bf Measurable} analogs of the {\bf momentum projection operator}
and the {\bf momentum angular projection operator} are studied.
\\In terms of the {\bf measurability} concept, analogs of
Fourier transformations are constructed. It is shown how to
produce a {\bf measurable} analog of the Heisenberg picture.
\\ At the end of the article, in Section 5, the obtained results are used
to substantiate another (more general) definition of the {\bf
measurability} concept that is not based on the Heisenberg
Uncertainty Principle and its generalizations, as it has been in
the earlier works of the author, and may be equally suitable both
for non-relativistic and relativistic cases.

\section{Measurability Concept}

\subsection{Primary Measurability in Nonrelativistic Case. Brief Consideration of the Principal Assumptions}

In this Subsection we briefly recall the principal assumptions
\cite{Shalyt-Entropy2016.1},\cite{Shalyt-ASTP3},\cite{Shalyt-new1}
that underlie further research.
\\
\\According to {\bf Definition I.} from
\cite{Shalyt-Entropy2016.1},\cite{Shalyt-ASTP3},\cite{Shalyt-new1}
we call as {\bf primarily measurable variation} any small
variation (increment) $\widetilde{\Delta} x_{i}$ of any spatial
coordinate $x_{i}$ of the arbitrary point $x_{i},i~=~1,...,3$ in
some space-time system $\emph{R}$ if it may be realized in the
form of the uncertainty (standard deviation) $\Delta x_{i}$ when
this coordinate is measured within the scope of Heisenberg's
Uncertainty Principle (HUP)\cite{Heis1},\cite{Mess}.
\\Similarly, we   call any small variation (increment) $\widetilde{\Delta}
x_{0}=\widetilde{\Delta} t_{0}$ by {\bf primarily measurable
variation}  in the value of time if it may be realized in the form
of the uncertainty (standard deviation) $\Delta x_{0}=\Delta t$
for pair ``time-energy'' $(t,E)$ when time is measured within the
scope of Heisenberg's Uncertainty Principle (HUP) too.
\\Next we introduce the following assumption:
\\{\bf Supposition II.} There is the minimal length
$l_{min}$  as {\it a minimal measurement unit} for all {\bf
primarily measurable variations} having the dimension of length,
whereas the minimal time $t_{min}=l_{min}/c$ as {\it a minimal
measurement unit} for all quantities or {\bf primarily measurable
variations (increments)}  having the dimension of time, where $c$
is the speed of light.
\\According to HUP $ l_ {min} $ and $ t_ {min} $ lead to $P_{max}$ and $E_{max}$.
For definiteness, we consider that $E_{max}$ and $P_{max}$ are the
quantities on the order of the Planck quantities, then $l_{min}$
and $t_{min}$ are also on the order of Planck quantities
$l_{min}\propto l_P$, $t_{min}\propto t_P$. {\bf Definition I.}
and {\bf Supposition II.} are quite natural in the sense that
there  are no physical principles with which they are
inconsistent.
\\The combination of {\bf Definition I.} and
{\bf Supposition II.} will be called the {\bf Principle of Bounded
Primarily Measurable Space-Time Variations (Increments)} or for
short {\bf Principle of Bounded  Space-Time Variations
(Increments)} with abbreviation (PBSTV).
\\As the minimal unit of measurement $l_{min}$ is available
 for all the {\bf primarily measurable variations} $\Delta L$
 having the dimensions of length, the
{``Integrality Condition'' (IC)} is the case
 \begin{equation}\label{Introd 2.4}
 \Delta L=N_{\Delta L}l_{min},
 \end{equation}
where~$N_{\Delta L}~>~0$~is an integer number.
\\In a like manner the same {``Integrality Condition'' (IC)}
is the case for all the {\bf primarily measurable variations}
$\Delta t$ having the dimensions of time. And similar to Equation
(\ref{Introd 2.4}), we get for any time $\Delta t$:
 \begin{equation}\label{Introd 2.4new}
\Delta t\equiv \Delta t(N_{t})=N_{\Delta t}t_{min},
\end{equation}
where similarly $N_{\Delta t}~>~0$ is an integer number too.
\\
\\{\bf Definition 1}({\textbf{Primary or Elementary
Measurability.}})
\\(1) {\it In accordance with PBSTV, let us define
the quantity having the dimensions of length or time  as {\bf
primarily (or elementarily) measurable}, when it satisfies the
relation Equation (\ref{Introd 2.4}) (and respectively Equation
(\ref{Introd 2.4new})}).
\\(2){\it Let us define any physical quantity {\bf
primarily (or elementarily) measurable},
 when its value is consistent with points (1)  of this
 Definition.}
\\
\\Here HUP is given for the nonrelativistic case.
In the next subsection we consider the relativistic case for low
energies $E\ll E_P$  and show that for this case {\bf Definition
1} of the ({\textbf{Primary Measurability}) retains its meaning. Further
for convenience, we everywhere denote the minimal length
$l_{min}\neq 0$ by $\ell$ and $t_{min}\neq 0$ by $\tau=\ell/c$.

\subsection{Primary Measurability in Relativistic Case}

In the Relativistic case HUP has its distinctive features
(\cite{Land2},Introduction). As known, in the relativistic case
for {\bf low energies} $E\ll E_P$, when the total energy of a
particle with the mass   $m$ and with the momentum $p$ equals
\cite{Land1}:
\begin{equation}\label{GUP1.new6}
E=\sqrt {{p^{2}}c^{2}+m^{2}c^{4}},
\end{equation}
a minimal value of $\Delta x$ in the general case takes the form
(\cite{Land2},formula(1.3))
\begin{equation}\label{GUP1.new7}
\Delta q\approx \frac{c \hbar}{E}=\frac{\hbar}{\sqrt
{{p^{2}}+m^{2}c^{2}}}.
\end{equation}
In this case, as well, nothing prevents the existence of the
minimal length $\ell\neq 0$ and the minimal time $\tau=\ell/c$ or
execution of the conditions (\ref{Introd 2.4}) and (\ref{Introd
2.4new}). Particularly, in the equation (\ref{Introd 2.4}) for
$\Delta L=\Delta q$, due to the fact that $E\ll E_P$, we have
 \begin{equation}\label{Introd 2.4R}
 \Delta q=N_{\Delta q}\ell;N_{\Delta q}\gg 1.
 \end{equation}
 The formula (\ref{GUP1.new7}) may be rewritten as
\begin{equation}\label{GUP1.new7R}
E\approx \frac{c \hbar}{N_{\Delta q}\ell}.
\end{equation}
Because $N_{\Delta q}\gg 1$ is an integer,
in the general case the energy $E$  may vary almost continuously, similarly to the canonical theory with $\ell=0$. A similar equation
(\ref{GUP1.new7R}) in this case may be derived for the momentum
$p$ from the right side of (\ref{GUP1.new7}) too.
It is clear that in the general case $p$ is also varying steadily.
\\An analogue of (\ref{GUP1.new7R}) is easily obtained in the
ultrarelativistic case ($E\approx p$) and in the rest frame of the
particle mass ($E\approx mc^{2}$). It is obvious that, according
to the above-mentioned equations, at {\bf low energies} the
picture is practically continuous.
\\ In the relativistic case, at least at   {\bf low energies} $E\ll
E_P$, {\bf Definition 1} of the ({\textbf{Primary Measurability})
from the previous subsection is also meaningful, though within the framework of the {\bf Uncertainty Principle for a Relativistic  System}
(\cite{Land2},Introduction).

\section{Generalized Measurability}

\subsection{Generalized Measurability and Generalized Uncertainty Principle}

Basic results  of  this  Subsection are given in
\cite{Shalyt-Entropy2016.1} and \cite{Shalyt-new1}.
\\Further it is convenient to use the deformation parameter
$\alpha_{a}$.
 This parameter has been introduced earlier in the papers
\cite{shalyt2},\cite{shalyt3},\cite{shalyt4}--\cite{shalyt7} as a
{\it deformation parameter} (in terms of paper \cite{Fadd}) on
going from the canonical quantum mechanics to  the quantum
mechanics at Planck's scales (Early Universe) that is considered
to be the quantum mechanics with the minimal length (QMML):
\begin{equation}\label{D1}
\alpha_{a}=\ell^{2}/a^{2},
\end{equation}
where $a$ is the primarily measuring scale. It is easily seen that
the parameter $\alpha_{a}$  from Equation (\ref{D1}) is discrete
as it is nothing else but
\begin{equation}\label{D1.new1}
\alpha_{a}=\ell^{2}/a^{2}=\frac{\ell^{2}}{N^{2}_{a}\ell^{2}}=\frac{1}{N^{2}_{a}}.
\end{equation}
for primarily measurable $a=N_{a}\ell$.
\\At the same time, from
Equation (\ref{D1.new1}) it is evident that $\alpha_{a}$ is
irregularly discrete.
\\It should be noted that, physical quantities complying with {\bf Definition 1}
 won't be enough for the research of physical systems.
 \\Indeed, such a variable as
\begin{eqnarray}\label{Def2}
\alpha_{N_{a}\ell}(N_{a}\ell)=p(N_{a})\frac{\ell}{\hbar}
=\ell/N_{a},
\end{eqnarray}
(where $\alpha_{N_{a}\ell}=\alpha_{a}$ is taken from formula
(\ref{D1.new1}) at $a=N_{a}\ell$, and
$p(N_{a})=\frac{\hbar}{N_{a}\ell}$ is  the corresponding {\bf
primarily  measurable} momentum), is fully expressed in terms {\it
only} {\bf Primarily Measurable Quantities} of {\bf Definition 1}
and that's why it may appear at any stage of calculations, but
apparently
 doesn't comply with {\bf Definition 1}.
 That's why it's necessary to introduce the following definition generalizing
{\bf Definition 1}:
\\
\\{\bf Definition 2. Generalized Measurability}
\\We shall call any physical quantity as {\bf
generalized-measurable}  or  for  simplicity {\bf measurable} if
any of its values may be obtained in terms of {\bf Primarily
Measurable Quantities} of {\bf Definition 1}.
\\
\\In what follows, for simplicity, we will use the term {\bf
 Measurability} instead of {\bf Generalized Measurability}.
It is evident that any {\bf primarily measurable quantity (PMQ)}
is {\bf measurable}. Generally speaking, the contrary is not
correct, as indicated by formula (\ref{Def2}).
\\It should be noted that Heisenberg's Uncertainty Principle (HUP)
\cite{Mess} is fair  at  low energies $E\ll E_P$. However it was
shown that at the Planck  scale a high-energy term must appear:
\begin{equation}\label{U2}
\Delta x\geq\frac{\hbar}{\Delta p}+ \alpha^{\prime}
l_{p}^2\frac{\triangle p}{\hbar}
\end{equation}
where $l_{p}$ is the Planck length $l_{p}^2 = G\hbar /c^3 \simeq
1,6\;10^{-35}m$ and $\alpha^{\prime}$ is a constant. In
\cite{Ven1} this term is derived from the string theory, in
\cite{GUPg1}
 it follows from the simple estimates of Newtonian gravity and quantum mechanics,
 in \cite{Magg2} it comes from the black hole physics, other methods can also be
 used \cite{Magg1},\cite{Magg3},\cite{Capoz}.
Relation (\ref{U2}) is quadratic in $\Delta p$
\begin{equation}\label{U4}
\alpha^{\prime} l_{p}^2\, ({\Delta p})^2 - \hbar\,\Delta x\Delta p
+ \hbar^2 \leq0
\end{equation}
 and therefore leads to the minimal length
\begin{equation}\label{U5}
 \Delta x_{min}=2 \surd \alpha^{\prime} l_{p}\doteq \ell
\end{equation}
Inequality (\ref{U2}) is called the Generalized Uncertainty
Principle (GUP) in Quantum Theory.
\\Let us show that the {\bf generalized-measurable} quantities  are appeared from the
{\bf Generalized Uncertainty Principle (GUP)}
\cite{Ven1}--\cite{Nozari} (formula (\ref{U2})) that naturally
leads to the minimal length $\ell$ (\ref{U5}).
\\Really solving inequality (\ref{U2}), in the case of equality we obtain the
apparent formula
\begin{equation}\label{root1}
\Delta p_{\pm}=\frac{(\Delta x\pm \sqrt{(\Delta
x)^{2}-4\alpha^{\prime}l^{2}_{p}})\hbar}{2\alpha^{\prime}l^{2}_{p}}.
\end{equation}
Next, into this formula we substitute the right-hand part of
formula (\ref{Introd 2.4}) for primarily measurable $L=x$.
Considering (\ref{U5}), we can derive the following:
\begin{eqnarray}\label{root2}
\Delta p_{\pm}=\frac{(N_{\Delta x}\pm \sqrt{(N_{\Delta
x})^{2}-1})\hbar\ell}{\frac{1}{2}\ell^{2}}=\nonumber\\
=\frac{2(N_{\Delta x}\pm \sqrt{(N_{\Delta x})^{2}-1})\hbar}{\ell}.
\end{eqnarray}
But it is evident that at low energies $E\ll E_p;N_{\Delta x}\gg
1$ the plus sign  in the nominator (\ref{root2}) leads to the
contradiction as   it results in very high (much greater than the
Planck’s) values of $\Delta p$. Because of this, it is necessary
to select the minus sign in the numerator (\ref{root2}). Then,
multiplying the left and right sides of (\ref{root2}) by the same
number $N_{\Delta x}+ \sqrt{N_{\Delta x}^{2}-1}$ , we get
\begin{eqnarray}\label{root3}
\Delta p=\frac{2\hbar}{(N_{\Delta x}+ \sqrt{N_{\Delta
x}^{2}-1})\ell}.
\end{eqnarray}
$\Delta p$ from formula (\ref{root3}) is the {\bf
generalized-measurable} quantity in the sense of {\bf Definition
2.} However, it is clear that at low energies $E\ll E_p$, i.e. for
$N_{\Delta x}\gg 1$, we have $\sqrt{N_{\Delta x}^{2}-1}\approx
N_{\Delta x}$. Moreover, we have
\begin{eqnarray}\label{root3.1}
\lim\limits_{N_{\Delta x}\rightarrow \infty}\sqrt{N_{\Delta
x}^{2}-1}=N_{\Delta x}.
\end{eqnarray}
Therefore, in this case (\ref{root3}) may be written as follows:
\begin{eqnarray}\label{root3.2}
\Delta p\doteq \Delta p(N_{\Delta x},
HUP)=\frac{\hbar}{1/2(N_{\Delta x}+ \sqrt{N_{\Delta
x}^{2}-1})\ell}\approx \frac{\hbar}{N_{\Delta
x}\ell}=\frac{\hbar}{\Delta x};N_{\Delta x}\gg 1,
\end{eqnarray}
in complete conformity with HUP. Besides, $\Delta p\doteq
\Delta p(N_{\Delta x},HUP)$, to a high accuracy,
 is a {\textbf{primarily measurable} quantity in the sense of {\bf
Definition 1}.
\\And vice versa it is obvious that at high energies $E\approx E_p$,
i.e. for $N_{\Delta x}\approx 1$, there is no way to transform
formula  (\ref{root3}) and we can write
\begin{eqnarray}\label{root3.3}
\Delta p\doteq \Delta p(N_{\Delta x},
GUP)=\frac{\hbar}{1/2(N_{\Delta x}+ \sqrt{N_{\Delta
x}^{2}-1})\ell};N_{\Delta x}\approx 1.
\end{eqnarray}
At the same time, $\Delta p\doteq \Delta p(N_{\Delta x},GUP)$ is a
{\textbf{Generalized Measurable} quantity in the sense of {\bf
Definition 2}.
\\ Thus, we have
\begin{equation}\label{root4}
GUP\rightarrow HUP
\end{equation}
for
\begin{equation}\label{root5}
 (N_{\Delta x}\approx 1)\rightarrow (N_{\Delta x}\gg 1).
\end{equation}
Also, we have
\begin{equation}\label{root6}
\Delta p(N_{\Delta x}, GUP)\rightarrow \Delta p(N_{\Delta x},HUP),
\end{equation}
where $\Delta p(N_{\Delta x}, GUP)$ is taken from formula
(\ref{root3.3}), whereas $\Delta p(N_{\Delta x}, HUP)$ from formula
 (\ref{root3.2}).
\\
\\{\it Comment 2*.}
\\{\it From the above formulae it follows that, within GUP, the {\bf primarily measurable}
variations (quantities) are derived to a high accuracy from the
{\bf generalized-measurable} variations (quantities) {\it only} in
the low-energy limit $E\ll E_P$}
\\
\\ Next, within the scope of GUP, we can correct a value of the parameter
$\alpha_{a}$  from formula  (\ref{D1.new1}) substituting $a$ for
$\Delta x$    in the expression $1/2(N_{\Delta x}+ \sqrt{N_{\Delta
x}^{2}-1})\ell$.
\\Then at low energies $E\ll E_p$ we have the
{\textbf{primarily measurable} quantity $\alpha_{a}(HUP)$
\begin{eqnarray}\label{root3.2.}
\alpha_{a}\doteq \alpha_{a}(HUP)=\frac{1}{[1/2(N_{a}+
\sqrt{N_{a}^{2}-1})]^{2}}\approx \frac{1}{N^{2}_{a}};N_{a}\gg 1,
\end{eqnarray}
that corresponds, to a high accuracy, to the value from formula (\ref{D1.new1}).
\\ Accordingly, at high energies we have
$E\approx E_p$
\begin{eqnarray}\label{root3.3.}
\alpha_{a}\doteq \alpha_{a}(GUP)=\frac{1}{[1/2(N_{a}+
\sqrt{N_{a}^{2}-1})]^{2}};N_{a}\approx 1.
\end{eqnarray}
When going from high energies $E\approx E_p$ to low energies
$E\ll E_p$, we can write
\begin{eqnarray}\label{P3}
\alpha_{a}(GUP) \stackrel{(N_{a}\approx 1)\rightarrow (N_{a}\gg
1)}{\longrightarrow} \alpha_{a}(HUP)
\end{eqnarray}
in complete conformity to {\it Comment 2*.}
\\{\bf Remark 3.1} What is the main difference between {\bf Primarily Measurable Quantities (PMQ)} and {\bf
Generalized Measurable Quantities (GMQ)}? {\bf PMQ}  defines
variables which may be obtained as a result of an immediate
experiment. {\bf GMQ}  defines the variables which may be  {\it
calculated} based on {\bf PMQ}, i.e. based on the data obtained in
previous clause.
\\
\\{\bf Remark 3.2.} It is readily seen that a minimal value of   $N_a=1$
is {\it unattainable} because in formula (\ref{root3.3}) we can
obtain a value of the length $l$ that is below the minimum
$l<\ell$ for the momenta and energies above the maximal ones, and
that is impossible. Thus, we always have $N_a\geq 2$. This fact
was indicated in \cite{shalyt2},\cite{shalyt3}, however, based on
the other approach.
\\
\\ As follows from the above formula, the {\bf generalized measurable}
 momenta at  all  energies  are of the form
\begin{eqnarray}\label{P3.new1}
p_{1/N}\doteq p(1/N,\ell),N\neq 0,
\end{eqnarray}
where $\ell=\kappa l_p$  up to the constant  $\kappa$ on the order of
1.
\\Therefore, $p_{1/N}$  depends only on three fundamental
constants $c,\hbar,G$, the constant $\kappa$, and the discrete parameter
$1/N$.
\\ But for $N\gg 1$, i.e. at $E\ll E_p$, the mapping of
$\tau:1/N \Rightarrow p_{1/N}$  is actually continuous providing
a high accuracy of coincidence between this discrete model and
the initial continuous theory.
\\The main target of the author is to resolve a quantum theory
and gravity in terms of the concept of {\bf primarily measurable}
quantities. As in this case these theories become discrete,
in what follows we use the lattice representation.

\subsection{Space and Momentum Lattices of Generalized Measurable Quantities,
 and  $\alpha-lattice$}

In this subsection the results from
\cite{Shalyt-Entropy2016.1},\cite{Shalyt-ASTP3}are refined and supplemented.
\\So, provided the minimal length $\ell$ exists, two lattices are
naturally arising by virtue of the formulas from the previous
subsection.
\\{\bf I.} At low energies (LE) $E\ll
E_{max}\propto E_P$, lattice of the space
variation---$Lat_{S}[LE]$ representing, for sets integers
$|N_{w}|\gg 1$  to within the known multiplicative constants, in
accordance with the above formulas for each of the three space
variables $w\doteq x;y;z$.

\begin{equation}\label{Latt2}
Lat_{S}[LE]= (N_{w}\doteq \{N_{x},N_{y},N_{z}\}),|N_{x}|\gg
1,|N_{y}|\gg 1,|N_{z}|\gg 1.
\end{equation}

At high energies (HE) $E\rightarrow E_{max}\propto E_P$ to within
the known multiplicative constants too in accordance with the
formulas previous subsection we  have the lattice $Lat_{S}[HE]$
for each of the three space variables $w\doteq x;y;z$.
\begin{equation}\label{Latt3}
Lat_{S}[HE]\doteq (\pm 1/2[(N_{w}+ \sqrt{N_{w}^{2}-1})]);2\leq
(N_{w}\doteq \{N_{x},N_{y},N_{z}\})\approx 1.
\end{equation}

{\bf II.} Next, let us define the lattice momentum variation
$Lat_{P}$
 as a set to {obtain} $(p_{x},p_{y},p_{z})$
for low energies $E\ll E_P$, where all the components of the above
sets conform to the space coordinates $(x,y,z)$, given by the
corresponding formulae  from the previous subsection.
\\From this it is inferred that, in analogy with point
I of this subsection, within the known multiplicative constants,
we have lattice $Lat_{P}[LE]$
\begin{equation}\label{Latt1.1}
Lat_{P}[LE]\doteq (\frac{1}{N_{w}}),
\end{equation}
where $N_{w}$ are~integer numbers from Equation (\ref{Latt2}).
\\In accordance with formulas (\ref{root3.3}), (\ref{Latt3}), the high-energy
(HE) momentum lattice $Lat_{P}[HE]$  takes the form
\begin{equation}\label{Latt3.1}
Lat_{P}[HE] \doteq(\pm\frac{1}{1/2[(N_{w}+ \sqrt{N_{w}^{2}-1})]}),
\end{equation}
where $N_{w}$ are~integer numbers from Equation (\ref{Latt3}).
\\It is important to note the following.

{\bf In the low-energy lattice $Lat_{P}[LE]$ all elements are
varying very smoothly, enabling the approximation of a continuous
theory}.
\\It is clear that  lattices $Lat_{S}[LE]$ and $Lat_{P}[LE]$ are
lattices of {\bf primarily measurable} quantities, while the lattices
$Lat_{S}[HE]$ and $Lat_{P}[HE]$ are lattices of the {\bf generalized
measurable} quantities.
\\
\\We will expand the space lattice  $Lat_{S}[LE]$ to space-time
lattice  $Lat_{S-T}[LE]$:
\begin{eqnarray}\label{Latt2.S}
Lat_{S-T}[LE]\doteq (N_{w},N_{t}),N_{w}\doteq
\{N_{x},N_{y},N_{z}\}, \nonumber
\\|N_{x}|\gg 1,|N_{y}|\gg 1,|N_{z}|\gg 1,|N_{t}|\gg 1
\end{eqnarray}
Now  {\bf primarily} lattice $Lat_{S-T}[LE]$ will be replaced with
$``\alpha-lattice``$, {\bf measurable space-time quantities},
which
 will be denoted by $Lat^{\alpha}_{S-T}[LE]$:
\begin{eqnarray}\label{Latt1.new1}
Lat^{\alpha}_{S-T}[LE]\doteq
(\alpha_{N_{w}\ell}N_{w}\ell,\alpha_{N_{t}\tau}N_{t}\tau)=(\frac{\ell^2}{\hbar}p(N_{w}),\frac{\ell^2}{\hbar}p(N_{t}))
=(\frac{\ell}{N_{w}},\frac{\tau}{N_{t}}).
\end{eqnarray}
In the last formula by the variable $\alpha_{N_{t}\tau}$ we mean
the parameter  $\alpha$ corresponding to the length $(N_{t}\tau)
c$:
\begin{equation}\label{Latt1.new0}
\alpha_{N_{t}\tau}\doteq \alpha_{(N_{t}\tau) c}.
\end{equation}
And $p(N_{w})$ it  is taken from  formula (\ref{Def2}), where
$N_{t}$ corresponds  formula (\ref{Latt1.new1}).
 As low energies
$E\ll E_P$ are discussed, $\alpha_{N_{w}\ell}$  in this formula is
consistent with the corresponding parameter from formula
(\ref{root3.2.}):
\begin{equation}\label{Latt1.new1.1}
\alpha_{N_{w}\ell}=\alpha_{N_{w}\ell}(HUP)
\end{equation}
As it was mentioned in the previous section,  in the low-energy
$E\ll E_{max}\propto E_P$ all elements of sublattice
$Lat_{P-E}[LE]$ are varying very smoothly enabling the
approximation of a continuous theory.
\\It is similar to the low-energy part of the  $Lat^{\alpha}_{S-T}[LE]$
 of lattice $Lat^{\alpha}_{S-T}$ will vary very smoothly:
\begin{equation}\label{Latt2.new}
Lat^{\alpha}_{S-T}[LE]=
(\frac{\ell}{N_{w}},\frac{\tau}{N_{t}});|N_{x}|\gg 1,|N_{y}|\gg
1,|N_{z}|\gg 1,|N_{t}|\gg 1.
\end{equation}
In Section 5 of \cite{Shalyt-Entropy2016.1} three following cases
were selected:
\\
\\(a){\it ``Quantum Consideration, Low Energies''}:
\\$$1\ll|N_{w}|\leq \widetilde{\textbf{N}},1\ll|N_{t}|\leq \widehat{\textbf{N}}$$
\\
\\(b){\it ``Quantum Consideration, High
Energies''}:
\\$$|N_{w}|\approx 1,|N_{t}|\approx 1;$$
\\
\\(c){\it ``Classical Picture''}:
$$|N_{w}|\rightarrow \infty,|N_{t}|\rightarrow \infty.$$
\\
Here $\widetilde{\textbf{N}},\widehat{\textbf{N}}$ is a cutoff
parameters, defined  by the current task
\cite{Shalyt-Entropy2016.1} and corrected in  this paper.
\\Let us for three space coordinates
$x_{i};i=1,2,3$ we introduce the following notation:
\begin{eqnarray}\label{Class1.1}
\Delta(x_{i})\doteq \widetilde{\Delta}[\alpha_{N_{\Delta
x_{i}}}]=\alpha_{N_{\Delta x_{i}}\ell}(N_{\Delta x_{i}}\ell) =
\ell/N_{\Delta x_{i}};\nonumber
\\
\frac{\Delta_{N_{\Delta x_{i}}} [F(x_{i})]}{\Delta(x_{i})}
\equiv\frac{F(x_{i}+\Delta(x_{i}))-F(x_{i})}{\Delta(x_{i})},
\end{eqnarray}
where $F(x_{i})$ is {\bf ''measurable''} function, i.e function
represented in terms of {\bf measurable} quantities.
\\Then function $\Delta_{N_{\Delta x_{i}}}[F(x_{i})]/\Delta(x_{i})$ is {\bf ''measurable''}
function too.
\\It's evident that
\begin{eqnarray}\label{Class2.1}
\lim\limits_{|N_{\Delta x_{i}}|\rightarrow
\infty}\frac{\Delta_{N_{\Delta x_{i}}} [F(x_{i})]}{\Delta(x_{i})}
=\lim\limits_{\Delta(x_{i})\rightarrow 0} \frac{\Delta_{N_{\Delta
x_{i}}} [F(x_{i})]}{\Delta(x_{i})}=\frac{\partial F}{\partial
x_{i}}.
\end{eqnarray}
Thus, we can define a {\bf measurable} analog of a  vectorial
gradient ${\bf \nabla}$
\begin{eqnarray}\label{Class2.1new1}
{\bf \nabla_{N_{\Delta x_{i}}}}\equiv\{\frac{\Delta_{N_{\Delta
x_{i}}} }{\Delta(x_{i})}\}
\end{eqnarray}
and a {\bf measurable} analog of the Laplace operator
\begin{eqnarray}\label{Class2.1new2}
{\bf \Delta}_{(N_{\Delta x_{i}})}\equiv{\bf \nabla_{N_{\Delta
x_{i}}}}{\bf \nabla_{N_{\Delta
x_{i}}}}\equiv\sum_{i}\frac{\Delta^{2}_{N_{\Delta x_{i}}}
}{\Delta(x_{i})^{2}}
\end{eqnarray}
 Respectively, for time $x_{0}=t$ we have:
\begin{eqnarray}\label{Class1.2}
\Delta(t)\doteq \widetilde{\Delta}[\alpha_{N_{\Delta
t}}]=\alpha_{N_{\Delta t}\tau}(N_{\Delta t}\tau) = \tau/N_{\Delta
t};\nonumber
\\
\frac{\Delta_{N_{\Delta t}} [F(t)]}{\Delta(t)}
\equiv\frac{F(t+\Delta(t))-F(t)}{\Delta(t)},
\end{eqnarray}
then
\begin{eqnarray}\label{Class2.2}
\lim\limits_{|N_{\Delta t}|\rightarrow \infty}
\frac{\Delta_{N_{\Delta t}}
[F(t)]}{\Delta(t)}=\lim\limits_{\Delta(t)\rightarrow 0}
\frac{\Delta_{N_{\Delta t}}[F(t)]}{\Delta(t)}=\frac{dF}{dt}.
\end{eqnarray}
We shall designate for momenta $p_{i};i=1,2,3$
\begin{eqnarray}\label{Class3.1}
{\Delta p_{i}}=\frac{\hbar}{N_{\Delta x_{i}}\ell};\nonumber
\\
\frac{\Delta_{p_{i}} F(p_{i})}{\Delta p_{i}}
\equiv\frac{F(p_{i}+\Delta p_{i})-F(p_{i})}{\Delta p_{i}}=
\frac{F(p_{i}+\frac{\hbar}{N_{\Delta
x_{i}}\ell})-F(p_{i})}{\frac{\hbar}{N_{\Delta x_{i}}\ell}}.
\end{eqnarray}
From where similarly  (\ref{Class2.1}) we get
\begin{eqnarray}\label{Class3.2}
\lim\limits_{|N_{\Delta x_{i}}|\rightarrow \infty}
\frac{F(p_{i}+\Delta p_{i})-F(p_{i})}{\Delta p_{i}}=
\lim\limits_{|N_{\Delta x_{i}}|\rightarrow
\infty}\frac{F(p_{i}+\frac{\hbar}{N_{\Delta
x_{i}}\ell})-F(p_{i})}{\frac{\hbar}{N_{\Delta x_{i}}\ell}}
=\nonumber
\\=\lim\limits_{\Delta p_{i}\rightarrow 0}
\frac{F(p_{i}+\Delta p_{i})-F(p_{i})}{\Delta p_{i}} =
\frac{\partial F}{\partial p_{i}}.
\end{eqnarray}
Therefore, at low energies $E\ll E_P$, i.e. at $|N_{\Delta
x_{i}}|\gg 1;|N_{\Delta t}|\gg 1, i=1,...,3$  on passage to the
limit (\ref{Class2.1}),(\ref{Class2.2}),(\ref{Class3.2}) we can
obtain from {\bf ''measurable''} functions the partial
derivatives like in the case of continuous space-time. That is, the
partial derivatives of {\bf ''measurable''} functions can be
considered as {\bf ''measurable''} functions with any given
precision.
\\In this case the infinitesimal space-time variations
(\ref{Introd.1}) are appearing in the limit from {\bf measurable}
quantities too
\begin{eqnarray}\label{Class7.4.new*}
(\alpha_{N_{\Delta t}\tau}N_{\Delta t}\tau=\frac{\tau}{N_{\Delta
t}}=p_{N_{\Delta t}c}\frac{\ell^{2}}{c\hbar}) \stackrel{N_{\Delta
t}\rightarrow \infty}{\longrightarrow}dt,\nonumber
\\(\alpha_{N_{\Delta x_{i}}\ell}N_{\Delta x_{i}}\ell=\frac{\ell}{N_{\Delta x_{i}}}=p_{N_{\Delta x_{i}}}\frac{\ell^{2}}{\hbar}) \stackrel{N_{\Delta x_{i}}\rightarrow
\infty}{\longrightarrow}dx_{i},1=1,...,3.
\end{eqnarray}
\\
\\{\bf Remark 3.2.1}
\\ As mentioned above, we suppose that the energies   $E$
are low, i.e. $E\ll E_p$.
\\ So far it has been connived that
all the numbers $N_{\Delta x_{i}},N_{\Delta t}$ are integers giving rise to the {\bf primarily measurable}  space-time
quantities $N_{\Delta x_{i}}\ell$ and $N_{\Delta t}\tau$.
Now this restriction is lifted because, unless it is specially noted otherwise, we assume that
$N_{\Delta x_{i}}\ell,N_{\Delta t}\tau$ are {\bf generalized
measurable} (or simply {\bf measurable}) quantities. At that, due
to the fact that the energies $E$ are low $E\ll E_P$, the following
condition is still true:
\begin{eqnarray}\label{Rem1}
|N_{\Delta x_{i}}|\gg 1;|N_{\Delta t}|\gg 1, i=1,...,3.
\end{eqnarray}
\\Therefore, in formula (\ref{Class7.4.new*}) the momenta $p_{N_{\Delta x_{i}}},p_{N_{\Delta
t}c}$ from this point onwards are {\bf generalized measurable}
quantities. Evidently, a good example of such momenta is an exact
rather than approximate value of the quantity from equation
(\ref{root3.2})
\begin{eqnarray}\label{root3.2*}
p_{N_{\Delta x_{i}}}=\frac{\hbar}{1/2(N_{\Delta x_{i}}+
\sqrt{N_{\Delta x_{i}}^{2}-1})}\ell;N_{\Delta x_{i}}\gg 1.
\end{eqnarray}
Besides, if $N_{\Delta x_{i}}\ell$ and $N_{\Delta
t}\tau$  are {\bf measurable} quantities, then the numeric
coefficients $N_{\Delta x_{i}}$ and $N_{\Delta t}$ are also {\bf
measurable} quantities.
\\In this case any {\bf measurable} triplet $N_{q}=\{N_{\Delta
x_{i}}\},|N_{\Delta x_{i}}|\gg 1,i=1,...,3$ corresponds to the small
{\bf measurable} momentum ${\bf p_{N_{q}}}\doteq \{p_{N_{\Delta
x_{i}}}\},$ with the components $p_{N_{\Delta x_{i}}},|p_{N_{\Delta
x_{i}}}|\ll P_{pl}$:
\begin{eqnarray}\label{Rem2}
N_{\Delta x_{i}}\stackrel{{\bf p}}{\longrightarrow}p_{N_{\Delta
x_{i}}}=\frac{\hbar}{ N_{\Delta x_{i}}\ell}.
\end{eqnarray}
And, vice versa, any small {\bf measurable} momentum ${\bf p_{q}}$
with the nonzero components ${\bf p_{q}}=\{p_{i}\};0\neq|p_{i}|\ll
P_{pl}$ corresponds to the {\bf measurable} triplet $N_{q}=\{N_{\Delta
x_{i}}\},|N_{\Delta x_{i}}|\gg 1,i=1,...,3$, satisfying the
condition (\ref{Rem1}):
\begin{eqnarray}\label{Rem3}
p_{i}\stackrel{{\bf x}}{\longrightarrow}N_{\Delta
x_{i}}=\frac{\hbar}{p_{i}\ell}.
\end{eqnarray}
For simplicity, instead of $N_{\Delta x_{\mu}}$,  we
use $N_{x_{\mu}},\mu=0,...,3$.

\section{Quantum Mechanics in Terms of Measurable Quantities}

\subsection{General Remarks on Wavefunction Representation}

Now any coordinate $u$ from the set $q\doteq (x,y,z)\in
 R^{3}$ and some {\bf measurable} quantity
$N_{u}\ell;|N_{u}|\gg 1$ we can correlate with the {\bf measurable}
quantity $\Delta_{N_{u}}(u)=\ell/N_{u}$, and
$N_{q}\doteq\{N_{x},N_{y},N_{z}\}$ -- with the {\bf measurable} product
\begin{eqnarray}\label{QM-1}
\Delta_{N_{q}}(q)=|\Delta_{N_{x}}(x)\cdot\Delta_{N_{y}}(y)\cdot\Delta_{N_{z}}(z)|=\frac{\ell^{3}}{|N_{x}N_{y}N_{z}|}.
\end{eqnarray}
Then it is clear that, for {\bf measurability} of the wave
function, $\Psi(q)$, ($\Psi(q)$  is determined in terms of the
{\bf measurability} concept of the spatial coordinates $q$, (i.e.
all variations of $q$ are {\bf measurable}). We can find the
quantity
\begin{eqnarray}\label{QM-2}
|\Psi(q)|^{2}\Delta_{N_{q}}(q)
\end{eqnarray}
which is the probability that the measurement performed for the
system will give the coordinate value in the {\bf measurable} volume
element $\Delta_{N_{q}}(q)$ of the configuration space.
\\ Then the known condition for the total probability in a continuous case \cite{Mess}
\begin{eqnarray}\label{QM-3.1}
\int|\Psi(q)|^{2}dq=1
\end{eqnarray}
is replaced, with any preassigned accuracy,  by the condition
\begin{eqnarray}\label{QM-3.2}
\sum_{q}|\Psi(q)|^{2}\Delta_{N_{q}}(q)=1.
\end{eqnarray}
Actually, due to equation (\ref{Class7.4.new*}), the {\bf
measurable} volume element $\Delta_{N_{q}}(q)$ of the
configuration space may be considered as arbitrarily close to  $dq$,
meaning that the {\bf measurable} element $q+\Delta_{N_{q}}(q)$
may be considered arbitrarily close to the {\bf nonmeasurable}  element $q+dq$.
\\It is obvious that a set of {\bf measurable} functions
forms the space, where the integrals of a continuous theory, if
any, are replaced by the corresponding sums  over the {\bf
measurable} quantities, and  $dq$ is replaced by
$\Delta_{N_{q}}(q)$. In the limit of high $|N_{q}|$, this space is
arbitrarily close to the corresponding Hilbert space of a
continuous theory.
\\In particular, the normalization condition for the {\bf measurable}
eigenfunction $\Psi_{n}$ of the given {\bf measurable} physical
quantity $f$  is varying from continuous to the {\bf
measurable} representation in the following way:
\begin{eqnarray}\label{QM-4}
(\int|\Psi_{n}|^{2}dq=1)\mapsto
(\sum_{q}|\Psi_{n}|^{2}\Delta_{N_{q}}(q)=1).
\end{eqnarray}
Similarly, we have
\begin{eqnarray}\label{QM-4.1}
\int\Psi\Psi^{\ast}dq\mapsto
\sum_{q}\Psi\Psi^{\ast}\Delta_{N_{q}}(q).
\end{eqnarray}
It is seen that, for the spaces of {\bf measurable} functions, we
can redefine all the principal properties of the canonical quantum
mechanics, superposition principle, properties of the operators
and of their spectra, replacing the integrals by the corresponding
sums and $dq$ -- by $\Delta_{N_{q}}(q)$ (as in the formula
(\ref{QM-4}),(\ref{QM-4.1})).

\subsection{Schrodinger Equation and Other Equations of Quantum Mechanics in ''Measurable'' Format}

\subsubsection{Schrodinger Equation for Free Particle}

Let us consider the Schrodinger Equation \cite{Mess} in terms of
{\bf measurable quantities}. As it is
shown in the formula (\ref{Class7.4.new*}) taking into account
{\bf Remark 3.2.1} at low energies $E\ll E_P$ (i.e. at
$|N_{x_{\mu}}|\gg 1$), the infinitesimal space-time variations
$dx_{\mu},\mu=0,...,3$ occur in the limit of $|N_{
x_{\mu}}|\rightarrow \infty$  from the {\bf measurable} momenta
$p_{N_{x_{i}}},(p_{N_{t}c})$ multiplied by the constant
$\frac{\ell^{2}}{\hbar},(\frac{\ell^{2}}{c\hbar})$ and are
nothing else but $\ell/N_{x_{i}},\tau/N_{t}$.
\\Therefore, in all the cases we assume that the following conditions are met:
 $|N_{x_{i}}|\gg 1,|N_{t}|\gg 1;i=1,...,3$.
\\Then a {\bf measurable} {\it $N_{t}$-analog} of the derivative
for the {\bf measurable} wave function $\Psi(t)$ in the continuous case
is nothing else but
\begin{eqnarray}\label{Schr-1.2}
\frac{\Delta_{N_{t}}[\Psi(t)]}{\Delta(t)}\doteq
\frac{\Psi(t+\tau/N_{t})-\Psi(t)}{\tau/N_{t}},
\end{eqnarray}
and a {\bf measurable} {\it $N_{t}$-analog} of the Schrodinger
Equation
\begin{eqnarray}\label{Schr-1.1}
\frac{d\Psi(t)}{dt}=\frac{1}{\imath\hbar}\widehat{H}\Psi(t)
\end{eqnarray}
is as follows:
\begin{eqnarray}\label{Schr-1.2*}
\frac{\Delta_{N_{t}}[\Psi(t)]}{\Delta(t)}=
\frac{\Psi(t+\tau/N_{t})-\Psi(t)}{\tau/N_{t}}=\frac{1}{\imath\hbar}\widehat{H}_{meas}\Psi(t).
\end{eqnarray}
Here $\widehat{H}_{meas}$ is some {\bf measurable} analog of the
Hamiltonian $\widehat{H}$ in the continuous case, i.e.,
$\widehat{H}_{meas}$ -- operator expressed in terms of {\bf
measurable} values.
\\Let us consider an example of the Schrodinger Equation for a free particle
\cite{Mess}
\begin{eqnarray}\label{Schr-new1}
\imath\hbar\frac{\partial}{\partial
t}\Psi(\textbf{r},t)=-\frac{\hbar^{2}}{2m}{\bf
\Delta}\Psi(\textbf{r},t),
\end{eqnarray}
where  ${\bf \Delta}\equiv{\bf \nabla}{\bf
\nabla}\equiv\frac{\partial^{2}}{\partial
x^{2}}+\frac{\partial^{2}}{\partial
y^{2}}+\frac{\partial^{2}}{\partial z^{2}}$ is the Laplace operator
and $m$ -- mass of the particle.
\\The formula (\ref{Class2.1new2}) has been initially considered for the case of integer
$N_{x_{i}},|N_{x_{i}}|\gg 1$. However, due to {\bf Remark 3.2.1},
it is still valid for any {\bf measurable} numbers
$N_{x_{i}},|N_{x_{i}}|\gg 1$.
\\From this formula it directly follows that
\begin{eqnarray}\label{Schr-new2}
\lim\limits_{|N_{x_{i}}|\rightarrow \infty}{\bf
\Delta}_{(N_{x_{i}})}={\bf \Delta}.
\end{eqnarray}
Then, because of the condition $|N_{t}|\gg 1,|N_{x_{i}}|\gg 1$, we
can infer that a {\bf measurable} analog of the Schrodinger
Equation (\ref{Schr-new1})
\begin{eqnarray}\label{Schr-new1*}
\imath\hbar\frac{\Delta_{N_{t}}}{\Delta(t)}\Psi(\textbf{r},t)=-\frac{\hbar^{2}}{2m}{\bf
\Delta}_{(N_{x_{i}})}\Psi(\textbf{r},t),
\end{eqnarray}
at rather high but finite $|N_{t}|,|N_{x_{i}}|$, is congruent with the
Schrodinger Equation in the continuous case to any preset accuracy.
\\Similary, from the formula for a {\bf measurable} value of the momentum at low energies $E\ll E_P$
\begin{eqnarray}\label{Schr-new3}
p_{N_{x_{i}}}=\frac{\hbar}{{N_{x_{i}}}\ell}
\end{eqnarray}
as well as the equation (\ref{Class2.1new1})  for a {\bf measurable}
analog of the vectorial gradient ${\bf \nabla_{N_{\Delta x_{i}}}}$
and the equations  (\ref{Class1.1}),(\ref{Class2.1})  it follows
that the correspondence rule in the {\bf measurable} case
\begin{eqnarray}\label{Schr-new3-M}
{\bf p_{N_{x_{i}}}}\doteq {\bf
p_{N_{q}}}\mapsto\frac{\hbar}{\imath}{\bf \nabla_{N_{q}}}
\end{eqnarray}
can, to any preset accuracy, reproduce the correspondence rule  in the
continuous case
\begin{eqnarray}\label{Schr-new3-C}
{\bf p}\mapsto\frac{\hbar}{\imath}{\bf \nabla}.
\end{eqnarray}
In a similar way for a {\bf measurable} value of the energy
\begin{eqnarray}\label{Schr-new4-M}
E_{N_{q}}=\frac{p^{2}_{N_{q}}}{2m}=\frac{p^{2}_{N_{x}}+p^{2}_{N_{y}}+p^{2}_{N_{z}}}{2m}
\end{eqnarray}
the correspondence
\begin{eqnarray}\label{Schr-new5-M}
E_{N_{q}}\mapsto\imath\hbar\frac{\Delta_{N_{t}}}{\Delta(t)}
\end{eqnarray}
reproduces the correspondence
\begin{eqnarray}\label{Schr-new5-M}
E\mapsto\imath\hbar\frac{\partial}{\partial t}
\end{eqnarray}
of a continuous theory.
\\So, in terms of {\bf measurable} quantities
 we can derive a discrete model arbitrary close the initial continuous theory.
\\From this it follows that the {\bf
measurable} wave function $\Psi_{meas}(\textbf{r},t,{\bf
N_{q}},N_{t})$ of the  form
\begin{eqnarray}\label{Wave1}
\Psi_{meas}(\textbf{r},t,{\bf
N_{q}},N_{t})=Aexp\{\imath(\frac{{\bf
p_{N_{q}}}\textbf{r}}{\hbar}-\frac{{\bf
E_{N_{q}}}\textbf{r}}{\hbar})  \},
\end{eqnarray}
where $\textbf{r}$  and $t$ -- {\bf measurable},
reproduces the corresponding wave function
$\Psi(\textbf{r},t)$ in the continuous case \cite{Mess} to a high accuracy.
\\ A particular example was given in preceding sections of the text.
It is obvious that it allows for more general conclusions.
\\ The {\bf measurable} analog $\widehat{H}_{meas}$  of the Hamiltonian $\widehat{H}$
from the equation (\ref{Schr-1.2*}) in the general case should be
of the following form:
\begin{eqnarray}\label{HAM-1}
\widehat{H}_{meas}=\widehat{H}_{meas}(N_{q}),
\end{eqnarray}
where $N_{q}$  is  {\bf measurable}
 and
\begin{eqnarray}\label{Schr-1.2**}
\lim\limits_{| N_{q}|\rightarrow
\infty}\widehat{H}_{meas}=\widehat{H}.
\end{eqnarray}
As we have
\begin{eqnarray}\label{Schr-1.3}
\lim\limits_{|N_{t}|\rightarrow
\infty}\frac{\Delta_{N_{t}}[\Psi(t))]}{\Delta(t)}=
\frac{d\Psi(t)}{dt},
\end{eqnarray}
then in the general case, in passage to the limit at
$|N_{q}|\rightarrow \infty,|N_{t}|\rightarrow \infty$, from a {\bf
measurable} analog of the Schrodinger equation (\ref{Schr-1.2*})
we can get the Schrodinger equation (\ref{Schr-1.1})  in the
continuous picture.
\\At that we can suppose that all variables, including time
$t$, influencing the wave function $\psi$ are {\bf measurable}
quantities. A similar supposition is correct for the Hamiltonian
$\widehat{H}_{meas}$ as well.
\\ Without loss of generality, we can assume that the values of
$| N_{q}|\gg 1$  are high enough so that the
{\bf measurable} Hamiltonian analog $\widehat{H}_{meas}$
be equal to the Hamiltonian in the continuous case to a high accuracy
\begin{eqnarray}\label{HAM-2}
\widehat{H}_{meas}=\widehat{H}
\end{eqnarray}
Then, at the fixed $N_{t}$, high in absolute value, and at {\bf
measurable} $\psi$, a {\bf measurable} analog of the Schrodinger
equation (\ref{Schr-1.2*}) may be solved recurrently as follows:
\begin{eqnarray}\label{Schr-1.2*1}
\Psi(t+\tau/N_{t})=(\frac{\tau}{\imath
N_{t}\hbar}\widehat{H}+1)\Psi(t).
\end{eqnarray}
Taking some {\bf measurable} quantity $\psi(t)$ (possibly $t=0$)
as a reference point and first substituting it into the right side
(\ref{Schr-1.2*1}), and then repeating this procedure for the
value of $\Psi(t+\tau/N_{t})$,  already calculated in the left
side, sufficiently many times, we can get the function
$\Psi(t+\Delta t)$  for arbitrary $\Delta t=K\tau/N_{t}$, where
$K$ is a natural number.
 It is clear that, if   $N_{t}$ -- integer number, then
{\bf primary measurable variations} in this series correspond to
only the integer $K$ divisible by $N_{t}$ , i. e.
$K=\mathcal{M}N_{t}$, where $\mathcal{M}$ -- integer number. But,
as the energies are low ($E\ll E_P$), we also have
$|\mathcal{M}|\gg 1.$.
\\ Then, because of
\begin{eqnarray}\label{Schr-1.4}
(\frac{\tau}{\imath N_{t}\hbar}\widehat{H}+1)\doteq
\widehat{U}(\tau/N_{t}),
\end{eqnarray}
we obtain
\begin{eqnarray}\label{Schr-1.4.1}
\frac{1}{\imath\hbar}\widehat{H}=\frac{\widehat{U}(\tau/N_{t})-1}{\tau/N_{t}}.
\end{eqnarray}
Next, assuming that $U(0)=1$ and considering
(\ref{Schr-1.2*})
\begin{eqnarray}\label{Schr-1.4*}
\frac{\Delta_{N_{t}}[\widehat{U}(t')]}{\Delta(t)}\doteq
\frac{\widehat{U}(t'+\tau/N_{t})-\widehat{U}(t')}{\tau/N_{t}},
\end{eqnarray}
we have
\begin{eqnarray}\label{Schr-1.5}
\frac{\Delta_{N_{t}}[\widehat{U}(t')]}{\Delta(t)}|_{t'=0}=\frac{1}{\imath\hbar}\widehat{H},
\end{eqnarray}
in strict conformity with the well-known formula in the continuous case
\begin{eqnarray}\label{Schr-1.6}
\widehat{H}=\imath\hbar
 \frac{d\widehat{U}(t')}{dt'}|_{t'=0}.
\end{eqnarray}
The operator $\widehat{U}(t')$ satisfying the equations
(\ref{Schr-1.4})--(\ref{Schr-1.5})  we denote as
$\widehat{U}_{N_{t}}$.
\\ From all the above formula it is trivial that
\begin{eqnarray}\label{Schr-1.7}
\Psi(t+\tau/N_{t})=\widehat{U}(\tau/N_{t})\Psi(t).
\end{eqnarray}
The presented calculations are easily generalized to non-autonomous
systems when the Hamiltonian $\widehat{H},(\widehat{H}_{meas})$
depends on time $t$, i.e. $\widehat{H}=\widehat{H}(t)$ and
the condition  (\ref{HAM-2}) is met. In this
case, again assuming that all values (operators and the wave function)
are {\bf measurable} quantities depending on time, we have
\begin{eqnarray}\label{Schr-1.8}
\Psi(t+\tau')=\widehat{U}(t+\tau',t)\Psi(t),\nonumber
\\
 \frac{\Delta_{N_{t}}}{\Delta(t)}\Psi(t)=\frac{\Delta_{N_{t}}[\widehat{U}(t+\tau',t)]}{\Delta(\tau')}|_{(\Delta(\tau')
 =\tau/N_{t})}\Psi(t)=\frac{1}{{\imath\hbar}}\widehat{H}(t)\Psi(t),\nonumber
\\ \widehat{H}(t)=\imath\hbar\frac{\Delta_{N_{t}}[\widehat{U}(t+\tau',t)]}{\Delta(\tau')}|_{\Delta(\tau').
 =\tau/N_{t}}
\end{eqnarray}
It is clear that in the suggested formalism one can reproduce all
the basic formulas of the continuous case replacing   $dt$ by
$\tau/N_{t}$, in particular
\begin{eqnarray}\label{Schr-1.9}
\widehat{U}^{\dagger}(t+\tau/N_{t},t)=(\widehat{1}+\frac{\tau}{N_{t}}\frac{\widehat{H}}{\imath
\hbar}+\textit{o}(\frac{\tau}{N_{t}}))^{\dagger}=\widehat{1}-\frac{\tau}{N_{t}}\frac{\widehat{H}^{\dagger}}{\imath
\hbar}+\textit{o}(\frac{\tau}{N_{t}})=\nonumber
\\=\widehat{U}^{-1}(t+\tau/N_{t},t)=(\widehat{1}+\frac{\tau}{N_{t}}\frac{\widehat{H}}{\imath
\hbar}+\textit{o}(\frac{\tau}{N_{t}}))^{-1}=\widehat{1}-\frac{\tau}{N_{t}}\frac{\widehat{H}}{\imath
\hbar}+\textit{o}(\frac{\tau}{N_{t}}).
\end{eqnarray}
What is the essence of replacing $dt$ by $\tau/N_{t}$ and of going
from continuous to the discrete picture in terms of {\bf
measurable quantities}? By the author’s opinion, the main point is
that the following {\bf Hypothesis} is valid:
\\
\\{\it at low energies $E\ll E_P$, i.e. at $|N_{t}|\gg 1$,
for any wave function $\Psi(t)$ there exists the integer ${\bf
N(\psi)},|{\bf N(\psi)}|\gg 1$  dependent on $\Psi(t)$,
with unimprovable approximation of the Schrodinger equation
(\ref{Schr-1.1}) by the discrete equation (\ref{Schr-1.2*}). Of
course, the condition $1\ll|N_{t}|\leq |{\bf N(\psi)}|$} is satisfied.

\subsubsection{Linear Momentum Operator}

It is known that a problem in eigenvalues and eigenfunctions of
the momentum projection $\hat{p}_{x_{i}}$  onto the coordinate
$x_{i}$ in the case of continuous space-time is reduced to the
differential equation \cite{Davydov}
\begin{eqnarray}\label{Mom-1.C}
-\imath\hbar\frac{\partial \Psi(x_{i})}{\partial
x_{i}}=p_{x_{i}}\Psi(x_{i}).
\end{eqnarray}
One can find continuous single-valued and bounded solutions of
this equation  for all real values of $p_{x_{i}}$ in the  interval
$-\infty<p_{x_{i}}<\infty$ with the eigenfunctions
\begin{eqnarray}\label{Mom-2.C}
\Psi_{p}(x_{i})=Aexp(\imath\frac{p}{\hbar}x_{i}).
\end{eqnarray}
Thus, we have one eigenfunction (no degeneracy) for each
eigenvalue $p_{x_{i}}=p$.
\\As stated above, in the {\bf measurable} case under consideration
in the left side  of (\ref{Mom-2.C}), for some  fixed {\bf
measurable} $|N_{x_{i}}|\gg 1$,  the replacement operation is used
\begin{eqnarray}\label{Mom-1.D}
\frac{\partial }{\partial x_{i}}\mapsto \frac{\Delta_{N_{x_{i}}}
}{\Delta(x_{i})}
\end{eqnarray}
and the eigenvalues $p_{N_{x_{i}}}$ of the operator
$\hat{p}_{x_{i}}$ become discrete $N_{x_{i}}$
\begin{eqnarray}\label{Mom-2.D}
p_{N_{x_{i}}}=\frac{\hbar}{N_{x_{i}}\ell},|N_{x_{i}}|\gg 1.
\end{eqnarray}
But, due to the condition $|N_{x_{i}}|\gg 1$,  we obtain a {\bf
discrete} spectrum of the operator $\hat{p}_{x_{i}}$ that is {\bf
nearly continuous}.
\\Taking into account that at sufficiently high $|N_{x_{i}}|$ ,
within any preset accuracy, we have
\begin{eqnarray}\label{Mom-1.Dnew}
\frac{\Delta_{N_{x_{i}}} }{\Delta(x_{i})}=\frac{\partial
}{\partial x_{i}}
\end{eqnarray}
and considering the formula (\ref{Mom-2.D}),  we can get
an analog of formula (\ref{Mom-2.C}) in the case under study
\begin{eqnarray}\label{Mom-2.Dnew}
\Psi_{p_{N_{x_{i}}}}(x_{i})=Aexp(\imath\frac{x_{i}}{N_{x_{i}}\ell}).
\end{eqnarray}
As seen, for fixed $x_{i}$,  the corresponding discrete
set of eigenfunctions also varies almost continuously.
\\It should be noted that the condition $-\infty<p_{x_{i}}<\infty$  in this case is incorrect because
\begin{eqnarray}\label{Mom-2.Dnew1}
((p_{x_{i}}=p_{N_{x_{i}}})\rightarrow \pm \infty)\equiv
(|N_{x_{i}}|\rightarrow 1),
\end{eqnarray}
that contradicts the condition $|N_{x_{i}}|\gg 1$.
\\However, for a real problem, the abstract condition $|N_{x_{i}}|\gg 1$
is always replaced by the specific condition
\begin{eqnarray}\label{N1}
|N_{x_{i}}|\geq {\bf N_{*}} \gg 1.
\end{eqnarray}
Then the condition $-\infty<p_{x_{i}}<\infty$  for the continuous case
 is replaced in the case under study by the condition $p_{-{\bf N_{*}}}\leq p_{x_{i}}\leq p_{{\bf N_{*}}}$,
with the distinguished point $p_{x_{i}}=0$ apparently not belonging to the equation (\ref{Mom-2.D}) at finite $N_{x_{i}}$.
\\It is clear that the case $N_{x_{i}}=\pm \infty$ associated with the point $p_{x_{i}}=0$
is degenerate and hence, if we limit ourselves to finite
$N_{x_{i}}$ , the condition  of (\ref{N1}) should be replaced with
the condition
\begin{eqnarray}\label{N2}
{\bf N^{*}}\geq|N_{x_{i}}|\geq {\bf N_{*}} \gg 1.
\end{eqnarray}
Then, respectively, we have $p_{x_{i}}\in [p_{-{\bf N_{*}}},p_{-{\bf N^{*}}}]\bigcup
[p_{{\bf N^{*}}}, p_{{\bf N_{*}}}]$.
\\Wext we denote with   $\Delta_{{\bf
 N_{*},N^{*}}}(p_{x_{i}})$ the integration of the intervals
 \begin{eqnarray}\label{N3}
\Delta_{{\bf
 N_{*},N^{*}}}(p_{x_{i}})\doteq [p_{-{\bf N_{*}}},p_{-{\bf
N^{*}}}]\bigcup [p_{{\bf N^{*}}}, p_{{\bf N_{*}}}],
\end{eqnarray}
and use $\Delta_{{\bf N_{*}}}({\bf p})$ for the following:
\begin{eqnarray}\label{N4}
\Delta_{{\bf N_{*},N^{*}}}({\bf p})=\prod_{i}\Delta_{{\bf
 N_{*},N^{*}}}(p_{x_{i}}).
\end{eqnarray}

\subsubsection {$z$-component of the Angular Momentum $\hat{L}_{z}$}

In the conventional quantum mechanics a problem in eigenvalues and
eigenfunctions of the angular momentum operator $\hat{L}_{z}$
\begin{eqnarray}\label{MAP-1}
\hat{L}_{z}=-\imath\hbar(x\frac{\partial }{\partial
y}-y\frac{\partial }{\partial x})
\end{eqnarray}
is reduced to solution of the  differential equation
\cite{Davydov}
\begin{eqnarray}\label{MAP-2}
-\imath\hbar\frac{\partial \Psi(\phi)}{\partial \phi}=
L_{z}\Psi(\phi),
\end{eqnarray}
where $0\leq\phi\leq 2\pi$.
\\In the considered case we can suppose that
$\phi=\phi(x,y,z)$--{\bf measurable} function of the variables
$x,y,z$ that in the continuous case has well-defined partial
derivatives for each of them.
\\It is  obvious  that  by substitution  into the formula  (\ref{Class2.1})  for $F(x_{i})=\phi(x,y,z)$  we immediately get
\begin{eqnarray}\label{MAP-3}
\lim\limits_{|N_{\Delta x_{i}}|\rightarrow
\infty}\frac{\Delta_{N_{\Delta x_{i}}}
[\phi(x,y,z)]}{\Delta(x_{i})}
=\lim\limits_{\Delta(x_{i})\rightarrow 0} \frac{\Delta_{N_{\Delta
x_{i}}} [\phi(x,y,z))]}{\Delta(x_{i})}=\frac{\partial
\phi}{\partial x_{i}}.
\end{eqnarray}
From whence it directly follows -- there exists the {\bf
measurable} function $\Delta\Psi/\Delta\phi$ so that
\begin{eqnarray}\label{MAP-4}
\lim\limits_{\Delta\phi\rightarrow 0}
\frac{\Delta\Psi}{\Delta\phi}=\lim\limits_{|N_{\Delta
x_{i}}|\rightarrow
\infty}\frac{\Delta\Psi}{\Delta\phi}=\frac{\partial \Psi}{\partial
\phi},
\end{eqnarray}
where  $\Delta \phi(x_{i})=\sum_{i}(\phi(x_{i}+\Delta
x_{i})-\phi(x_{i}))$  and {\bf measurable} increments of $\Delta
x_{i}$ are taken from the formula (\ref{Class1.1}).
\\ Considering that, for sufficiently high $|N_{x_{i}}|$,
with a high accuracy we have
\\$\Delta_{N_{x_{i}}}/\Delta(x_{i})=\partial/\partial
x_{i}$ and
$\Delta\Psi(\phi)/\Delta\phi=\partial\Psi(\phi)/\partial\phi$, it
is concluded that the equation (\ref{MAP-2}) to a high accuracy
may be used in the {\bf measurable} case as well, when regarding
$\phi(x,y,z))$ the {\bf measurable} function of a {\bf
measurable}set of the coordinates $\{x,y,z\}$.
\\Then a solution for (\ref{MAP-2}) is given by the exponent
\begin{eqnarray}\label{MAP-5}
\Psi(\phi)=Aexp(\imath\frac{L_{z}}{\hbar}\phi),
\end{eqnarray}
where   $\phi=\phi(x,y,z)$--{\bf measurable} function of the {\bf
measurable} variables $x,y,z$.
\\ In so doing the eigenfunctions for a discrete spectrum $L_{z}=\hbar m; m=0,\pm1,\pm2,...$
of the operator  $\hat{L}_{z}$, as in the continuous case, are given by
\begin{eqnarray}\label{MAP-6}
\Psi_{m}(\phi)
=(2\pi)^{-1/2}e^{\imath m\phi},
\end{eqnarray}
where $\phi$ is a  {\bf measurable}  quantity.
\\However, at the normalization condition the integral in the continuous case \cite{Davydov} is replaced by the sum
\begin{eqnarray}\label{QM-4.new1}
(\int_{0}^{2\pi}|\Psi_{m}|^{2}d\phi=1)\Rightarrow
(\sum_{0\leq\phi\leq 2\pi}|\Psi_{m}|^{2}\Delta(\phi)=1),
\end{eqnarray}
where $\Delta(\phi)$  is taken from the formula (\ref{MAP-4}).

\subsection{Position and Momentum Representations, and Fourier Transform in Terms of Measurability}

Now, using the formulas from the previous sections, let us study
quantum representations and the Fourier transform in terms of the
concept of {\bf measurable} quantities. A scalar product in the
position representation in the continuous case is given by the
equality \cite{Mess},\cite{Fadd2}:
\begin{equation} \label{Quant2}
(\varphi_{1},\varphi_{2})=\int_{R^{3}}\varphi^{*}_{1}({\bf
x})\varphi_{2}({\bf x})d{\bf x}.
\end{equation}
Both the coordinate ${\bf x_{j}}$  and momentum ${\bf
p_{j}}$,($j=1,2,3$) operators in the position representation are
introduced by \cite{Mess}
\begin{eqnarray}\label{Quant3}
{\bf x_{j}}.\varphi({\bf x})=x_{j}\varphi({\bf x}),\nonumber
\\{\bf p_{j}}.\varphi({\bf x})=-\imath\hbar\frac{\partial}{\partial x_{j}}\varphi({\bf x}).
\end{eqnarray}
According to the formula (\ref{QM-1}), we have ${\bf x}=q$ and hence
the integral from the equation
(\ref{Quant3}) is replaced by the sum
\begin{equation} \label{Quant2.M}
(\varphi_{1},\varphi_{2})_{meas}=\sum_{{\bf x}\in R^{3}
}\varphi^{*}_{1}\varphi_{2}\Delta_{N_{{\bf x}}}({\bf x}),
\end{equation}
where ${\bf x}$ -- {\bf measurable} coordinates.
\\It is clear that the passage to the limit takes place
\begin{equation} \label{Quant2.M1}
\lim\limits_{N_{x_{i}}\rightarrow
\infty}(\varphi_{1},\varphi_{2})_{meas}=(\varphi_{1},\varphi_{2}),
\end{equation}
where $\{N_{x_{i}}\}=N_{q}$ due to the equation (\ref{QM-1}) and at
sufficiently high  $\{N_{x_{i}}\}=N_{q}$ with a high precision we have
\begin{equation} \label{Quant2.M2}
(\varphi_{1},\varphi_{2})_{meas}=(\varphi_{1},\varphi_{2}).
\end{equation}
In the considered case the first formula from (\ref{Quant3}) is
valid for all {\bf measurable} values in the left and right
sides, while the second one is replaced by
\begin{eqnarray}\label{Quant3.1}
{\bf p_{N_{x_{j}}}}.\varphi({\bf
x})=-\imath\hbar\frac{\Delta_{N_{x_{j}}}}{\Delta(x_{j})}\varphi({\bf
x})\doteq \nonumber
\\ \doteq-\imath\hbar\frac{\varphi(x_{i\neq j}, x_{j}+\ell/N_{x_{j}})-\varphi({\bf
x})}{\ell/N_{x_{j}}}.
\end{eqnarray}
Here ${\bf p_{N_{x_{j}}}}$ -- j-th {\bf measurable} component of the momentum taking the form

\begin{eqnarray}\label{Quant3.1P}
p_{N_{x_{j}}}=\frac{\hbar}{N_{x_{j}}\ell}.
\end{eqnarray}
and the function $\varphi(x_{i\neq j}, x_{j}+\ell/N_{x_{j}})$
differs from $\varphi({\bf x})$ only by the “shift” to
$\ell/N_{x_{j}}$ in the j-component.
\\As follows from the formulae given above and, in particular, the formula (\ref{Class2.1}), in this low-energy case when
$E\ll E_P$, i.e. at $|N_{x_{j}}|\gg 1$ , to a high accuracy we
have
\begin{eqnarray}\label{Quant3.1.new}
\frac{\Delta_{N_{x_{j}}}}{\Delta(x_{j})}=\frac{\partial}{\partial
x_{j}}.
\end{eqnarray}
Then, due to the formulae (\ref{Quant3.1})--(\ref{Quant3.1.new}),
 in the low-energy case $E\ll E_P$ for
{\bf measurable} quantities within a high accuracy we get
\begin{eqnarray} \label{UR1.new3**}
[{\bf{x}},{\bf{p}}].\varphi({\bf x})
={\bf{x}}{\bf{p}}.\varphi({\bf x})-{\bf{p}}{\bf{x}}.\varphi({\bf
x})=\imath\hbar\varphi({\bf x}).
\end{eqnarray}
In the momentum representation for the continuous picture we have
\begin{eqnarray}\label{Quant3.P}
{\bf x_{j}}.\varphi({\bf p})=\imath\hbar\frac{\partial}{\partial
p_{j}}\varphi({\bf p}),\nonumber
\\{\bf p_{j}}.\varphi({\bf p})=p_{j}\varphi({\bf p}).
\end{eqnarray}
In the {\bf measurable} case the second equation (\ref{Quant3.P})
for {\bf measurable} momenta remains unchanged.
 According to the formulae
(\ref{Class3.1}) and (\ref{Class3.2}), in the {\bf measurable} case in the first equation from
(\ref{Quant3.P}) the replacement is performed
\begin{eqnarray}\label{Quant4.P}
\frac{\partial}{\partial p_{j}}\mapsto \frac{\Delta_{p_{j}}
}{\Delta p_{j}},
\end{eqnarray}
where
\begin{eqnarray}\label{Quant5.P}
p_{j}\doteq p_{N_{x_{j}}}=\frac{\hbar}{N_{x_{j}}\ell};\nonumber
\\
\frac{\Delta_{p_{j}} \varphi({\bf p}))}{\Delta p_{j}}
\equiv\frac{\varphi({\bf p}+p_{j})-\varphi({\bf p})}{p_{j}}=
\frac{\varphi({\bf p}+\frac{\hbar}{N_{ x_{j}}\ell})-\varphi({\bf
p})}{\frac{\hbar}{N_{x_{j}}\ell}},
\end{eqnarray}
and $\varphi({\bf p}+p_{j})$  differs from $\varphi({\bf p})$ by $p_{j}$  only in  the  j-th component.
\\Then from the expression
(\ref{Class3.2}), due to the fact that $|N_{x_{j}}|\gg 1$, with a high
accuracy we get
\begin{eqnarray}\label{Quant4.PE}
\frac{\Delta_{p_{j}} }{\Delta p_{j}}=\frac{\partial}{\partial
p_{j}}.
\end{eqnarray}
\\Now let us consider $[{\bf{x}},{\bf{p}}].\varphi({\bf p})$
in the momentum representation. Taking into account the formula
(\ref{Quant4.PE}), we have
\begin{eqnarray} \label{UR1.new2}
[{\bf{x}}_{j},{\bf{p}}_{j}].\varphi({\bf p})\
={\bf{x}}_{j}{\bf{p}}_{j}.\varphi({\bf
p})-{\bf{p}}_{j}{\bf{x}}_{j}.\varphi({\bf p})=\nonumber
\\=\imath \hbar
(\varphi({\bf p})+p_{j}\frac{\varphi({\bf p}+p_{j})-\varphi({\bf
p})}{p_{j}}-\nonumber
\\
-p_{j}\frac{\varphi({\bf p}+p_{j})-\varphi({\bf p}}{
p_{j}})=\nonumber
\\
= \imath\hbar .\varphi({\bf p}).
\end{eqnarray}
Thus, the expressions (\ref{Quant3.1.new})--(\ref{UR1.new2}) show
that the commutator
\begin{eqnarray} \label{COMM*}
[{\bf{x}}_{i},{\bf{p}}_{j}]=\imath\delta_{ij}\hbar
\end{eqnarray}
in the {\bf measurable} case occurs both in the position and
momentum representations.
\\In the continuous picture the Fourier transform
is of the following form \cite{Fadd2}:
\begin{eqnarray} \label{FUR1}
\varphi({\bf x})=(\frac{1}{2\pi
\hbar})^{3/2}\int_{R^{3}}e^{\frac{\imath}{\hbar}{\bf
px}}\varphi({\bf p})d{\bf p}.
\end{eqnarray}
And the operator ${\bf p_{j}}$  applied to the formula (\ref{FUR1})
gives \cite{Fadd2}
\begin{eqnarray} \label{FUR1.1}
{\bf p_{j}}.\varphi({\bf x})=-\imath\hbar\frac{\partial}{\partial
x_{j}}\varphi({\bf x})=-\imath\hbar\frac{\partial}{\partial
x_{j}}(\frac{1}{2\pi
\hbar})^{3/2}\int_{R^{3}}e^{\frac{\imath}{\hbar}{\bf
px}}\varphi({\bf p})d{\bf p}=\nonumber
\\=(\frac{1}{2\pi
\hbar})^{3/2}\int_{R^{3}}e^{\frac{\imath}{\hbar}{\bf
px}}p_{j}\varphi({\bf p})d{\bf p}.
\end{eqnarray}
However, as indicated in the formulae
(\ref{Mom-2.Dnew1}),(\ref{N1}), in the  considered {\bf
measurable} case at low energies the values of $|{\bf p}|$ are
bounded, therefore ${\bf p}$ fills not the whole space of $R^{3}$,
belonging only to its part $\Delta_{{\bf N_{*},N^{*}}}({\bf p})$
(formula (\ref{N4})).
\\That is why the integral in the equation
(\ref{FUR1}) should be replaced by the sum
\begin{eqnarray} \label{FUR1M}
\varphi_{meas}({\bf x})=(\frac{1}{2\pi \hbar})^{3/2}\sum_{{\bf
p}\in\Delta_{{\bf N_{*},N^{*}}}({\bf
p})}e^{\frac{\imath}{\hbar}{\bf px}}\varphi_{meas}({\bf
p})\Delta_{p}(p_{N_{\bf x}}),
\end{eqnarray}
where ${\bf x}$,${\bf p}$ and $\varphi_{meas}({\bf p})$ are {\bf
measurable} quantities. So, we have
\begin{eqnarray} \label{FUR2M}
\Delta_{p}(p_{N_{\bf x}})=\prod_{j}p_{N_{x_{j}}},
\end{eqnarray}
where $p_{N_{x_{j}}}$  is taken from the equation
(\ref{Quant5.P}).
\\As $|N_{x_{j}}|\gg 1$, then in the limit $|N_{x_{j}}|\rightarrow \infty$ the sum in the right side of the equation (\ref{FUR1M})
 is replaced by the integral and, to a high accuracy, we get \begin{eqnarray} \label{FUR1M-1}
(\frac{1}{2\pi \hbar})^{3/2}\int_{\Delta_{{\bf N_{*},N^{*}}}({\bf
p})}e^{\frac{\imath}{\hbar}{\bf px}}\varphi({\bf p})d{\bf p}=
\\ \nonumber =
(\frac{1}{2\pi \hbar})^{3/2}\sum_{{\bf p}\in\Delta_{{\bf
N_{*},N^{*}}}({\bf p})}e^{\frac{\imath}{\hbar}{\bf
px}}\varphi_{meas}({\bf p})\Delta_{p}(p_{N_{\bf x}}).
\end{eqnarray}
It should be noted that in this case the domain of the function
varies only for the momenta. Due to the above-mentioned equations,
it is narrowing
 $\{{\bf p}\in
R^{3}\}\mapsto \{{\bf p}\in\Delta_{{\bf N_{*},N^{*}}}({\bf p})\}$.
For the coordinates, it remains $\{{\bf x}\in R^{3}\}$.
\\The function $\varphi({\bf p})$  in the continuous case is of the following form \cite{Fadd2}:
\begin{eqnarray} \label{FUR1.P}
\varphi({\bf p})=(\frac{1}{2\pi
\hbar})^{3/2}\int_{R^{3}}e^{-\frac{\imath}{\hbar}{\bf
px}}\varphi({\bf x})d{\bf x}.
\end{eqnarray}
As the domain in the position representation remains the same
$\{{\bf x}\in R^{3}\}$, then for the {\bf measurable} case
$\varphi_{meas}({\bf p})$ takes the form
\begin{eqnarray} \label{FUR1.PM}
\varphi_{meas}({\bf p})=(\frac{1}{2\pi
\hbar})^{3/2}\sum_{R^{3}}e^{-\frac{\imath}{\hbar}{\bf
px}}\varphi_{meas}({\bf x})\Delta_{N_{\bf x}}({\bf x}),
\end{eqnarray}
where ${\bf x}=q$ from the formula (\ref{QM-1}), i.e.
\begin{eqnarray}\label{QM-1.FUR}
\Delta_{N_{{\bf x}}}({\bf
x})=\prod_{j}\Delta_{N_{x_{j}}}(x_{j})=\frac{\ell^{3}}{N_{x}N_{y}N_{z}}.
\end{eqnarray}
In this case, due to the condition $|N_{x_{j}}|\gg 1$, we have \begin{eqnarray}\label{FUR1M-2}
(\frac{1}{2\pi
\hbar})^{3/2}\int_{R^{3}}e^{-\frac{\imath}{\hbar}{\bf
px}}\varphi({\bf x})d{\bf x}\approx (\frac{1}{2\pi
\hbar})^{3/2}\sum_{R^{3}}e^{-\frac{\imath}{\hbar}{\bf
px}}\varphi_{meas}({\bf x})\Delta_{N_{\bf x}}({\bf x}),
\end{eqnarray}
where all values in the right side of (\ref{FUR1M-2})  are {\bf
measurable}.
\\Thus, the equations (\ref{FUR1M}) and (\ref{FUR1.PM} are
analogues of direct and inverse Fourier transforms in terms of
{\bf measurable} quantities or, better to say, of the {\bf measurable}
direct and inverse Fourier transforms.
\\In this formalism we can easily derive a {\bf measurable}
analog of the equation (\ref{FUR1.1})  by the replacement of ${\bf
p_{j}}\mapsto{\bf p_{N_{x_{j}}}}$,$\partial/\partial x_{j}\mapsto
\Delta_{N_{x_{j}}}/\Delta(x_{j})$,$\varphi({\bf
x})\mapsto\varphi_{meas}({\bf x})$ and $\int_{R^{3}}\mapsto
\sum_{\Delta_{{\bf N_{*}}}({\bf p})}.$
\\Similarly, by the adequate replacement in the {\bf measurable}
variant it is possible  to get  an analog of the correspondence
\begin{eqnarray}\label{FUR4}
{\bf x_{j}}.\varphi({\bf
p})\mapsto\imath\hbar\frac{\partial}{\partial p_{j}}\varphi({\bf
p})
\end{eqnarray}
in the continuous picture.
\\
\\It is necessary to make some important comments.
\\
\\{\bf Commentary 4.3.}
\\
\\{\bf 4.3.1.} As we consider the minimal length $\ell$ and the minimal time $\tau$
at Plank’s level $\ell\propto l_p,\tau\propto t_p$, the use of the
{\bf measurable} quantities $\ell/N_{x_{i}};i=1,...,3$ and
$\tau/N_{t}$ at $|N_{x_{i}}|\gg 1,|N_{t}|\gg 1$ as a substitution
for $dx_{i},dt$ in the continuous case is absolutely correct and
justified.   Indeed, as in this case $\ell$  is on the order of
$\approx 10^{-33}cm$, then $\ell/N_{x_{i}}$ is on the order of
$\approx 10^{-33-\lg|N_{x_{i}}|}cm$, without doubt being beyond
any computational accuracy. A similar statement is true for
$\tau/N_{t}$ as well, where $\tau$ is on the order of the Plank
time $t_p$, i.e. $\approx 10^{-44}sec$. For this reason, it is
correct to use in the continuous case $p_{N_{x_{i}}}$ instead of
$dp_{i}$ and
$\Delta_{N_{x_{i}}}/\Delta(x_{i}),\Delta_{N_{t}}/\Delta(t),\Delta_{p_{i}}/\Delta
p_{i}$ instead of $\partial/\partial x_{i},\partial/\partial
t,\partial/\partial p_{i}$, respectively.
\\
\\{\bf 4.3.2.} For generality, in {\bf Remark 3.2.1} we have supposed
that $N_{x_{i}},N_{t}$ are the {\bf generalized measurable}
quantities. As $|N_{x_{i}}|\gg 1,|N_{t}|\gg 1$, without loss of
generality, we can regard $N_{x_{i}}$ and $N_{t}$  as {\bf
primarily  measurable} quantities. It is clear that
\begin{eqnarray}\label{Comm 4.3-1}
[N_{x_{i}}]\leq N_{x_{i}}\leq [N_{x_{i}}]+1,
\end{eqnarray}
where $[\aleph]$ defines the integer part of $\aleph$. Then
$|N_{x_{i}}|^{-1}$ falls within the interval with the finite
points $|[N_{x_{i}}]|^{-1}$ and $|[N_{x_{i}}]+1|^{-1}$ (which of
the numbers is greater or smaller, depends on a sign of
$N_{x_{i}}$). In any case we have
$|N_{x_{i}}^{-1}-[N_{x_{i}}]^{-1}|\leq
|([N_{x_{i}}]+1)^{-1}-[N_{x_{i}}]^{-1}|=|([N_{x_{i}}]+1)[N_{x_{i}}]|^{-1}$.
\\ Actually, the difference between $\ell/N_{x_{i}}$ and
$\ell/[N_{x_{i}}]$ (or, respectively, between
$\Delta_{N_{x_{i}}}/\Delta(x_{i})$ and
$\Delta_{[N_{x_{i}}]}/\Delta(x_{i})$, and  so on) is
insignificant. Similar computations are correct for $\tau/N_{t}$
and $\tau/[N_{t}]$ as well.
\\
\\{\bf 4.3.3a.}It is important: despite the fact that in the {\bf measurable} case
we have analogues of the direct and inverse Fourier transforms
specified by the equations (\ref{FUR1M})  and (\ref{FUR1.PM}), the
difference between the position and momentum representations in
this case is significant. Indeed, for the first the domain is
represented by the whole three-dimensional space $R^{3}$, whereas
for the second the domain represents a particular part of the
finite dimensions $\Delta_{{\bf N_{*},N^{*}}}({\bf p})$, ''cut
out''  from the three-dimensional space $\Delta_{{\bf
N_{*},N^{*}}}({\bf p})\subset R^{3}$.
\\
\\{\bf 4.3.3b.} A significant difference between the position and momentum representations
in the {\bf measurable} case is associated with their different
nature in this formalism. In principle, the position
representation in this case is formed similar to the continuous
case. The momentum representation in the {\bf measurable} case, as
follows from the formulas {\bf Remark 3.2.1}, is formed on the
basis of {\bf measurable variations}   in the coordinate
representation.
\\As, to within a multiplicative constant, $\ell$ agrees
with $l_p$ and $p_{N_{\bf x}}$ -- with $\ell/N_{\bf x}$ (formula
(\ref{Class7.4.new*})), the measures on summation in the {\bf
measurable} case in equations (\ref{FUR1M}) and (\ref{FUR1.PM})
for the momentum and position spaces also agree to within the
multiplicative constant
\begin{eqnarray}\label{Measure 1}
\Delta_{N_{{\bf x}}}({\bf
x})=\frac{\ell^{6}}{\hbar^{3}}\Delta_{p}(p_{N_{\bf x}}).
\end{eqnarray}
\\{\bf 4.3.4.} Note that the above-mentioned formalism
used to study the Schrodinger picture in terms of {\bf
measurability} may be applied to the Heisenberg picture too
\cite{Mess},\cite{Fadd2}. Indeed, in the paradigm of the
continuous space and time a motion equation  for the Heisenberg
operators $\hat{L(t)}$ is as follows \cite{Mess},\cite{Fadd2}:
\begin{eqnarray}\label{Heis 1}
\frac{d\hat{L}(t)}{dt}=\frac{\partial\hat{L}(t)}{\partial
t}+[\hat{H},\hat{L}(t)],
\end{eqnarray}
where $\hat{H}$ -- Hamiltonian  and
$[\hat{H},\hat{L}(t)]=\frac{1}{\imath
\hbar}(\hat{L}(t)\hat{H}-\hat{H},\hat{L}(t))$--quantum Poisson bracket  \cite{Fadd2}.
\\In the {\bf measurable} case the quantum Poisson bracket preserves its form for the enclosed {\bf measurable} quantities.
$\partial\hat{L}(t)/\partial t$  is replaced by
$\Delta_{N_{t}}[\hat{L}(t)]/\Delta(t)$, where the operator
$\Delta_{N_{t}}[\hat{L}(t)]/\Delta(t)$ may be obtained from
equation (\ref{Schr-1.4*}) due to replacement of $\widehat{U}(t')$
by $\hat{L}(t)$ at $|N_{t}|\gg 1.$
\\Then an analogue of (\ref{Heis 1}) in the {\bf measurable} case is given by
\begin{eqnarray}\label{Heis 1M}
\frac{\tilde{\Delta}_{N_{t}}[\hat{L}(t)]}{\Delta(t)}\doteq\frac{\Delta_{N_{t}}[\hat{L}(t)]}{\Delta(t)}+[\hat{H},\hat{L}(t)].
\end{eqnarray}
It is clear that
\begin{eqnarray}\label{Heis 2}
\lim\limits_{|N_{t}|\rightarrow
\infty}\frac{\tilde{\Delta}_{N_{t}}[\hat{L}(t)]}{\Delta(t)}=\frac{d\hat{L}(t)}{dt}.
\end{eqnarray}
Thus, at sufficiently high $|N_{t}|$,  equation (\ref{Heis 1M})
agrees with equation (\ref{Heis 1}) to a high accuracy.

\section{More General Definition of Measurability}

Proceeding from all the above, the author suggests another
definition of {\bf measurability} that is more general
than the initial one.
\\ As before, we begin with a particular minimal (universal)
unit for measurement of the length  $\ell$ corresponding to some
maximal energy $E_\ell=\frac{\hbar c }{\ell}$ and a universal unit
for measurement of time $\tau=\ell/c$. Without loss of generality,
we can consider $\ell$ and $\tau$ at Plank’s level, i.e.
$\ell=\kappa l_p,\tau=\kappa t_p$, where the numerical constant
$\kappa$ is on the order of 1. Consequently, we have
$E_\ell\propto E_p$ with the corresponding   proportionality
factor.
\\ Note that $\ell$ and $\tau$ are referred to as ''minimal'' and ''universal''
units of measurement because in our case this is actually true.
\\Now consider in the space of momenta ${\bf
P}$ the domain defined by the conditions
\begin{eqnarray}\label{Meas-D1}
{\bf p}=\{p_{x_i}\},i=1,..,3;P_{pl}\gg|p_{x_i}|\neq 0,
\end{eqnarray}
where $ P_{pl}$--Plank momentum. Then we can easily calculate the
numerical coefficients $N_{x_i}$ as follows:
\begin{eqnarray}\label{Meas-D2}
N_{x_i}=\frac{\hbar}{p_{x_i}\ell}, or
\\ \nonumber p_{x_i}\doteq p_{N_{x_i}}=\frac{\hbar}{N_{x_i}\ell}
\\ \nonumber|N_{x_i}|\gg 1,
\end{eqnarray}
where the last row of the equation (\ref{Meas-D2}) is given by
the formula (\ref{Meas-D1}).
\\
\\{\bf Definition 1*}
\\{\bf 1*.1} The momenta ${\bf p}$ given by the formula   (\ref{Meas-D1})
are called {\bf primarily measurable} when all $N_{x_i}$  from the
equation (\ref{Meas-D2}) are integer numbers.
\\{\bf 1*.2}. Any variation in $\Delta x_i$ for the coordinates $x_i$
and $\Delta t$ of the time $t$ at the energies  $E\ll E_p$
is considered {\bf primarily measurable} if
\begin{eqnarray}\label{Meas-D3}
\Delta x_i=N_{x_i}\ell,\Delta t=N_{t}\tau,
\end{eqnarray}
where $N_{x_i}$  satisfies the condition  {\bf 1*.1} and
$|N_{t}|\gg 1$ -- integer number.
\\{\bf 1*.3} Let us define any physical quantity as {\bf
primary or elementary measurable} at  low energies $E\ll E_p$
when its value is consistent with points {\bf 1*.1} and {\bf 1*.2}
of this Definition.
\\
\\For convenience, we denote a domain of the momenta
satisfying the conditions (\ref{Meas-D1}) (or (\ref{Meas-D2}))  in terms of ${\it{\bf P}}_{LE}.$.
\\In Commentary {\bf 4.3.2} it is shown that, since the energies are low $E\ll E_P$ ($|N_{x_i}|\gg 1$), {\bf primary
measurable} momenta are sufficient to find the whole domain  of momenta ${\it{\bf P}}_{LE}.$
\\This means that in the indicated domain a discrete set of {\bf primary
measurable} momenta $p_{N_{x_i}};i=1,...,3$ (where
$N_{x_i}$--integer number and $|N_{x_i}|\gg 1$),  varies almost
continuously, practically covering the whole domain.
\\That is why further ${\it{\bf P}}_{LE}$ is associated with the domain  of
 {\bf primary measurable} momenta, satisfying the conditions of
the formula (\ref{Meas-D1}) (or (\ref{Meas-D2})).
\\
\\Then boundaries of the domain ${\it{\bf
P}}_{LE}$ are determined by the condition (\ref{N2}) for each
coordinate
\\
$${\bf N^{*}}\geq|N_{x_{i}}|\geq {\bf N_{*}} \gg 1,$$,
\\
where high natural numbers ${\bf N^{*}},{\bf N_{*}}$ are
determined by the problem at hand.
\\The choice of the number ${\bf N^{*}}$ is of particular importance.
If ${\bf N^{*}}< \infty$, then it is clear that the studied
momenta fall within the domain ${\it{\bf P}}_{LE}$. Assuming the
condition ${\bf N^{*}}= \infty$, to  ${\it{\bf P}}_{LE}$ for every
coordinate $x_i$  we should add ''improper'' (or ''singular'')
point $p_{x_i}=0$ (these cases  are called {\bf degenerate}).
\\In any case, for each coordinate $x_i$, the boundaries of ${\it{\bf P}}_{LE}$ are of the form:
\begin{eqnarray}\label{Meas-D4old}
p_{\bf N^{*}}\leq|p_{N_{x_{i}}}|\leq p_{\bf N_{*}}
\end{eqnarray}
For definiteness, we denote ${\it{\bf P}}_{LE}$, having the
boundaries  specified by the formula (\ref{Meas-D4old}), in terms
of ${\it{\bf P}}_{LE}[{\bf N^{*}},{\bf N_{*}}].$
\\It is obvious that in this formalism {\bf small}
increments for any component $p_{N_{x_i}}$  of the momentum ${\bf
p}\in {\it{\bf P}}_{LE}$ are values of the momentum
$p_{N^{'}_{x_i}}$, so that $|N^{'}_{x_i}|>|N_{x_i}|$. And then,
incrementing $|N^{'}_{x_i}|$, we can obtain  {\bf arbitrary small}
increments for the momenta ${\bf p}\in {\it{\bf P}}_{LE}$.
\\ In this case it is correct to define a “measurable partial derivative”
with respect to the momentum $p_{N_{x_i}}$  specified in equations (\ref{Class3.1}) and (\ref{Class3.2})
in terms of $\frac{\Delta_{p_{N_{x_i}}}}{\Delta p_{N_{x_i}}}$.
 As shown by the equations (\ref{Class3.1}) and (\ref{Class3.2})
and in the previous paragraph, at sufficiently high $|N_{x_i}|$,  with any predetermined accuracy,  we have the equality
$\frac{\Delta_{p_{N_{x_i}}}}{\Delta
p_{N_{x_i}}}=\frac{\partial}{\partial p_i}$  (for
example, formula (\ref{Quant4.PE})).
\\ It is clear that {\bf primary measurable} variations in $\Delta x_i$
for the coordinates $x_i$ and in $\Delta t$ for the time $t$ given
in point {\bf 1*.2} of {\bf Definition 1*} could hardly play a
role of small spatial and temporal variations.   Still, based on
the equation (\ref{Class7.4.new*}) and its applications in
subsequent parts of the text, we can state that the space and time
quantities
\begin{eqnarray}\label{Meas-D4}
\frac{\tau}{N_{t}}=p_{N_{t}c}\frac{\ell^{2}}{c\hbar} \nonumber
\\
\frac{\ell}{N_{x_{i}}}=p_{N_{x_{i}}}\frac{\ell^{2}}{\hbar},1=1,...,3
\end{eqnarray}
are small and, see (\ref{Class7.4.new*}),they may be arbitrary
small for sufficiently high values of $|N_{x_{i}}|,|N_{t}|$. Here
$p_{N_{x_{i}}},p_{N_{t}c}$--corresponding {\bf primarily
measurable} momenta.
\\ Of course, due to point 1.2 of Definition 1,
the space and time quantities
$\frac{\tau}{N_{t}},\frac{\ell}{N_{x_{i}}}$ are not {\bf primary
measurable} despite the fact that they, to within a constant
factor, are equal to {\bf primarily measurable} momenta.
\\Therefore, it seems expedient to introduce the following definition:
\\
\\{\bf Definition 2*.(Generalized Measurability at Low Energies).}
\\ Any physical quantity at low energies $E\ll E_p$ may be called
{\bf generalized measurable} or, for simplicity, {\bf measurable}
if any of its values may be obtained in terms of {\bf Primary
Measurable Quantities} specified by {\bf Definition 1*}.
\\
\\Lifting the restriction of $P_{pl}\gg|p_{x_i}|$  in the equation (\ref{Meas-D1}) or,
similarly, of $|N_{x_i}|\gg 1$ in the formula (\ref{Meas-D2}),
i.e. considering the momenta space ${\bf p}$ at {\bf all energy
scales}
\begin{eqnarray}\label{Meas-D*}
{\bf p}=\{p_{x_i}\},i=1,..,3;|p_{x_i}|\neq 0;\\ \nonumber
N_{x_i}=\frac{\hbar}{p_{x_i}\ell}, or
\\ \nonumber p_{x_i}\doteq p_{N_{x_i}}=\frac{\hbar}{N_{x_i}\ell},
\\ \nonumber 1\leq|N_{x_i}|<\infty,or \quad  E\leq E_\ell,
\end{eqnarray}
we can  introduce the following definition.
\\
\\{\bf Definition 3*(Primary and Generalized Measurability at All Energy Scales).}
\\{\bf 3*.1.} The momenta ${\bf p}$, set by the formula (\ref{Meas-D*}),
are referred to as {\bf primarily measurable}, if all the numbers
$N_{x_i}$  from this formula  (\ref{Meas-D*}) are integers.
\\{\bf 3*.2.} Any variation $\Delta x_i$ in the coordinates $x_i$ and $\Delta t$
in the time $t$ at all energy scales $E\leq E_\ell$ are referred to as {\bf primary measurable} if
\begin{eqnarray}\label{Meas-D3*}
\Delta x_i=N_{x_i}\ell,\Delta t=N_{t}\tau,
\end{eqnarray}
where $N_{x_i}$  satisfies the condition {\bf 3*.1} and the integer
$N_{t}$  falls within the interval $1\leq|N_{t}|<\infty$.
\\{\bf 3*.3.} We define any physical quantity as {\bf
primary or elementary measurable} at  all energies scales
$E\leq E_\ell$ when its value is consistent with points {\bf
3*.1} and {\bf 3*.2} of this Definition.
\\{\bf 3*.4.} Finally, we define any physical quantity at all energy scales $E\leq E_\ell$
as {\bf generalized measurable} or, for simplicity, {\bf
measurable} if any of its values may be obtained in terms of {\bf
Primarily Measurable Quantities} specified by points  {\bf
3*.1}--{\bf 3*.3} of {\bf Definition 3*}.
\\
\\ The ''improper'' points associated with $|N_{x_i}|=\infty$ and  $|N_{t}|=\infty$
may be introduced into equation(\ref{Meas-D*}) and into {\bf
Definition 3*}, respectively, as in the case of low energies.
\\ It has been shown that the {\bf Primary Measurable Momenta}
nearly cover the whole momenta domain  ${\it{\bf P}}_{LE}$ at low
energies $E\ll E_p$ (or identically $E\ll E_\ell$). However, this is no
longer the case  at {\bf all the energy scales} $E\leq E_\ell$.
\\Therefore, the main target of the author is to construct
a quantum theory at all energy scales $E\leq E_\ell$ in terms  of
{\bf measurable} (or identically {\bf primary measurable})
quantities from {\bf Definition 3*}.
\\In this theory the values of the physical quantity $\mathcal{G}$
may be represented by the numerical function $\mathcal{F}$ in the following way:
\begin{eqnarray}\label{Meas-D*F}
\mathcal{G}=\mathcal{F}(N_{x_i},N_{t},\ell)=\mathcal{F}(N_{x_i},N_{t},G,\hbar,c,\kappa),
\end{eqnarray}
where $N_{x_i},N_{t}$--integers from the formulae
(\ref{Meas-D*}),(\ref{Meas-D3*}) and $G,\hbar,c$ are  fundamental
constants. The last equality in (\ref{Meas-D*F}) is determined by
the fact that  $\ell=\kappa l_p$ and $l_p=\sqrt{G\hbar/c^{3}}$.
\\If $N_{x_i}\neq 0,N_{t}\neq 0$ ({\bf nondegenerate} case),
then it is clear that  (\ref{Meas-D*F}) can be rewritten as follows:
\begin{eqnarray}\label{Meas-D*F*}
\mathcal{G}=\mathcal{F}(N_{x_i},N_{t},\ell)=\mathcal{\widetilde{F}}((N_{x_i})^{-1},(N_{t})^{-1},\ell)
\end{eqnarray}
Then at low energies $E\ll E_p$, i.e. at $|N_{x_i}|\gg
1,|N_{t}|\gg 1$, the function $\mathcal{\widetilde{F}}$ is a
function of the variables changing practically  continuously,
though these variables cover a discrete set of values. Naturally,
it is assumed that $\mathcal{\widetilde{F}}$ varies smoothly (i.e.
practically continuously). As a result, we get a model, discrete
in nature, capable to reproduce the well-known theory in
continuous space-time to a high accuracy, as it has been stated
above.
\\Obviously, at low energies $E\ll E_p$ the formula (\ref{Meas-D*F*})
is as follows:
\begin{eqnarray}\label{Meas-D*F*P}
\mathcal{G}=\mathcal{F}(N_{x_i},N_{t},\ell)=\mathcal{\widetilde{F}}((N_{x_i})^{-1},(N_{t})^{-1},\ell)=
\\ \nonumber
=\mathcal{\widetilde{F}}_p(p_{N_{x_i}},p_{N_{tc}},\ell),
\end{eqnarray}
where $p_{N_{x_i}},p_{N_{tc}}$ are {\bf primary measurable}
momenta from formula (\ref{Class7.4.new*}).
\\It should be noted that our approach to the concept {\bf
measurability}, as set forth in this Section, is considerably more general than in Sections 2,3 for two reasons:
\\a)it is not directly related to the Heisenberg Uncertainty Principle and its generalizations;
\\
\\b)it can be successfully used both for the nonrelativistic \cite{Mess} and relativistic cases \cite{Peskin}.

\section{Final Comments and Further Prospects}

{\bf 6.1.} Thus, for all the energy scales, we can derive a model
still to be constructed) that is dependent on the same discrete
parameters and, at low energies $E$ far from the Planck energies
$E\ll E_p$, is very
 close to the initial theory, reproducing all the main results
of the canonical quantum theory in continuous space-time to a high
accuracy. At high (Planck’s) energies $E\approx E_p$ this discrete
model is liable to give new results.
\\ By the author’s opinion, the model is deprived of the principal drawbacks
of the canonical quantum theory--ultraviolet and infrared
divergences \cite{Peskin}. Being finite at all orders of a
perturbation theory, it requires no renormalization
procedures\cite{Peskin}.
\\
\\{\bf 6.2.} As shown by the formula (\ref{Class7.4.new*}) and (\ref{Meas-D4}), {\bf measurable} analogs of small and infinitesimal
space-time quantities are equal (up to constants) to the {\bf
primary measurable} momenta.
\\This allows us to state for gravity \cite{Einst1} the same problem as for a quantum theory in paragraph {\bf 6.1.}:
\\to construct a {\bf measurable} model of gravity
depending on the same discrete parameters $N_{x_i},N_{t}$, that is
practically continuous and ''very close'' to General Relativity at
low energies $E\ll E_p$,   giving the correct quantum theory
without ultraviolet divergences at high energies $E\approx
E_p,(E\approx E_\ell)$.
\\Hovewer, the words ''very close''
in the last paragraph don’t mean that there is an ideal
correspondence between the above-mentioned model and General
Relativity \cite{Einst1}. The author assumes that the desired
model should have no ''nonphysical'' solutions of General
Relativity (for example, those involving the {\bf Closed Time-like
Curves} (CTC) \cite{Godel}--\cite{Lobo}).
\\
\\{\bf 6.3.} At the present time each of the involved theories (Quantum Theory and Gravity)
considered within the scope of continuous space-time is
represented differently at low $E\ll E_p$ and at high $E\approx
E_p$ energies.
\\ We can summarize points  {\bf 6.1.} and {\bf 6.2.} as  follows.
\\
\\In the {\bf measurable} format, both these theories (quantum theory
and gravity) represent a unified theory at  all energies scales
$E\leq E_\ell$. The word ''unified'' means that at all the energy
scales they should be determined by the same discrete set of the
parameters $N_{x_i},N_{t}$ and  by the constants
$G,\hbar,c,\kappa$.
\\The main problem in this case is associated with a correct definition and computation of the functions
$\mathcal{F}$ and $\mathcal{\widetilde{F}}$ from formula
(\ref{Meas-D*F})--(\ref{Meas-D*F*P}).
\\In Subsection 3.1, within the scope of the Generalized Uncertainty Principle,
we have already found the function $\mathcal{F}$  for all {\bf
measurable} momenta  $p_{i,{\bf meas}};i=1,..,3$ at  {\bf all the
energy scales} $E\leq E_\ell$ by the formula
(\ref{root3.2}),(\ref{root3.3}):
\begin{eqnarray}\label{Meas-P}
p_{i,{\bf
meas}}=\mathcal{F}(N_{x_i},\ell)=\frac{\hbar}{1/2(N_{x_i}+
\sqrt{N_{x_i}^{2}-1})\ell}.
\end{eqnarray}

\begin{center}
{\bf Conflict of Interests}
\end{center}
The author declares that there is no conflict of interests
regarding the publication of this paper.
\\
\\{\bf Acknowledgment}
\\This work was supported by the Belarusian
Republication Foundation for Fundamental Research (project {\bf
N}16-121).

\end{document}